\documentclass[aps,prl,twocolumn,showpacs]{revtex4-1}
\usepackage{amsbsy,latexsym,amsmath,amsthm,amsfonts,mathrsfs,bbold}
\usepackage{amsfonts}
\usepackage{amssymb}
\usepackage{epsfig,graphics,graphicx}

\newcommand{\ket}[1]{\vert#1\rangle}
\newcommand{\bra}[1]{\langle#1\vert}
\newcommand{\mean}[1]{\langle#1\rangle}
\newcommand{\braket}[2]{\langle#1\vert#2\rangle}
\newcommand{\ketbra}[2]{\vert#1\rangle\langle#2\vert}
\newcommand{\ketbrad}[1]{\vert#1\rangle\langle#1\vert}

\newcommand{\id}{\openone}
\newcommand{\mse}{\sigma^{2}}
\newcommand{\msej}{\sigma^{2}_{j}}

\newcommand{\tr}[1]{\mathrm{tr}(#1)}

\newcommand{\expo}[1]{\mathrm{e}^{#1}}

\begin{document}
\title{Probabilistic metrology defeats ultimate deterministic bound}
\author{J. Calsamiglia, B. Gendra, R. Mu\~{n}oz-Tapia, and E. Bagan}
\address{F\'{i}sica Te\`{o}rica: Informaci\'{o} i Fen\`{o}mens Qu\`antics, Departament de F\'{\i}sica, Universitat Aut\`{o}noma de Barcelona, 08193 Bellaterra (Barcelona), Spain}

\begin{abstract}
Quantum-enhanced measurements exploit quantum mechanical effects to provide ultra-precise estimates of physical variables for use in advanced technologies, such as frequency calibration of atomic clocks, gravitational waves detection, and biosensing. Quantum metrology studies the fundamental limits in the estimation precision given a certain amount of resources (e.g. the number of probe systems) and restrictions (e.g. limited interaction time, or coping with unavoidable presence of noise). 
Here we show that, even in the presence of noise, probabilistic measurement strategies (which have a certain probability of failure or abstention) can provide, upon a heralded successful outcome, estimates with a precision that violates the deterministic bounds. This   establishes a new ultimate quantum metrology limit. For probe systems subject to local dephasing, we quantify such precision limit as a function of the probability of failure that can be tolerated. We show that the possibility of abstaining can substantially set back the detrimental effects of noise

\end{abstract}

\maketitle

Quantum-enhanced precision measurements and sensors are one the most disruptive quantum technologies \cite{dowling_quantum_2003} with applications  across various disciplines, e.g.  optical communications \cite{slavik_all-optical_2010,chen_optical_2012},  cryptography \cite{inoue_differential_2002},  brain and heart  medical diagnosis  via atomic magnetometry \cite{budker_optical_2007,napolitano_interaction-based_2011}, biological measurements \cite{taylor_biological_2013,crespi_measuring_2012}; and are critical in gravitational-wave detectors \cite{caves_quantum-mechanical_1981,collaboration_gravitational_2011} and GPS and other current technologies that rely on atomic clocks   \cite{bollinger_optimal_1996,huelga_improvement_1997,wineland_spin_1992,komar_quantum_2014}.

In broad terms a metrology problem  can be cast as a four step process: the preparation of a probe, its controlled interaction with the (continuous) parameter to be estimated,  the measurement of the modified probe and a final data-processing to infer the value of the unknown parameter.  The accuracy of the estimation is limited by the experimental imperfections and, ultimately, by the noise  inherent in any quantum measurements. Classically, the only way to reduce the effects of noise in a given setup is to repeat the experiment a number, $n$, of times \cite{giovannetti_quantum_2006}. The resulting precision of the estimation is thereby reduced by a factor $n^{{-1/2}}$ (the so-called standard quantum limit, SQL). However, in a fully quantum mechanical setting, the possibility of  using  entangling operations in the preparation and measurement steps gives rise to a  precision that scales as $n^{{-1}}$ (the so-called Heisenberg scaling).

Recent experimental advances that allow an unprecedented control of diverse optical and condensed matter  systems at a quantum level makes quantum metrology an extremely timely field of research \cite{giovannetti_advances_2011} . In the last years the agenda of quantum-enhanced metrology has been put under scrutiny by a number of results \cite{demkowicz-dobrzanski_elusive_2012,escher_general_2011,knysh_scaling_2011,aspachs_phase_2009,huelga_improvement_1997} that show that under  quite generic (local, uncorrelated and markovian) experimental noise, the  quantum enhancement in the asymptotic limit of infinite $n$ amounts  to a constant factor rather than quadratic improvement.  The field has revamped in search for alternative schemes that push forward the limits and circumvent or diminish the detrimental effect of noise. This has entailed the of study particular systems with non-trivial noise-models \cite{chaves_noisy_2013,chin_quantum_2012,jeske_quantum_2013,ostermann_protected_2013,szankowski_parameter_2012}, and  non-linear interactions \cite{boixo_generalized_2007,napolitano_interaction-based_2011}, which enable quantum error-correction codes  \cite{dur_improved_2014,kessler_quantum_2014,arrad_increasing_2014}.

In this work we put forward an extra feature that is always available to an experimentalist, namely the possibility of post-selecting the outcomes of their measurements.  Up until now most quantum metrology schemes and known bounds have been deterministic, that is they are optimized in order to provide a valid estimate for each possible measurement outcome.  Only recently,  it was shown that for a fixed probe state and in the absence of noise the precision of the favorable outcomes can be greatly enhanced well beyond the limits set for deterministic strategies \cite{fiurasek_optimal_2006,gendra_quantum_2013,gendra_optimal_2013,marek_optimal_2013}, of course at the price of discarding or abstaining on the unfavourable outcomes. The possibility of abstaining can even change the precision from SQL to Heisenberg scaling.  It has also been shown that the probabilistic  quantum limit agrees with that found for deterministic strategies when optimizing over probe preparations. So,  for pure states post-selection can compensate a bad choice of probe state, or in other words, it can attain optimal precision bounds in situations where the probe state is a given. 

Here, we show that in the presence of local dephasing, probabilistic metrology can lessen substantially the effects of noise, although not enough to overcome the infamous loss of asymptotic Heisenberg scaling \cite{escher_general_2011,demkowicz-dobrzanski_elusive_2012}. In addition, and in contrast to the noiseless ideal case, the ultimate precision bounds obtained exceed those that can be obtained by deterministic strategies --even when they are optimized over probe states. 

From a technical point of view we introduce a novel technique \footnote{The analogy with the quantum mechanical problem of the groundstate of a particle in a potential was first presented by J. Calsamiglia,  Abstention-enhanced metrology, Noise Information \& Complexity @ Quantum Scale, Ettore Majorana Centre, Erice, Italy (2013). A similar equivalence has been drawn recently in \cite{knysh_true_2014} in a point-wise approach (see Methods)} to compute the optimal probabilistic measurement and its precision by showing that it is formally equivalent to finding the the ground state and energy of a particle in a one-dimensional potential box, with some boundary conditions that depend on the strength of the noise and on the initial probe state.

Before moving to the detailed account of our results let us briefly discuss 
the use of probabilistic strategies in parameter estimation, and in general quantum information processing.
This is pertinent specially in view of some recent criticism on probabilistic metrology schemes \cite{combes_quantum_2014}. 
 As mentioned above there are tasks where it is critical to have an estimate of a given precision, for example timing the signals in a GPS setup or, probably less practical, hitting a small target with a billion dollar rocket. It is not hard to imagine that if the error in the estimated frequency or the estimated location of the target, respectively, exceeds a given critical value, then it can lead to catastrophic consequences or unsurmountable money losses. It is also perfectly plausible that, given a particular state (correlated or uncorrelated)  of $n$ subsystems\footnote{In our approach we take the total number of subsystems as a resource. This is in contrast to other (usually point-wise) approaches where the number of subsystems in a single run is accounted, but an unlimited number of repetitions of each run is allowed. See Methods.},  the precision bounds obtained for deterministic strategies can exceed that critical error value. Faced with this situation the experimentalist may choose to abort the task entrusted to her, say, the launch of the rocket. However, if she  decides to proceed with the experiment, it could well be that the error bar around  some particular outcome (or sequence of outcomes if one considers several consecutive measurements on the $n$ copies)  is below the critical value. That is, although the average over all measurement outcomes cannot exceed the deterministic bounds, some  measurement outcomes may produce better estimates than others. In particular, for some outcome the error may fall below the critical value, and the entrusted task can after all be carried out successfully. This type of error-assessment can be  done in a classical setting or for a fixed choice of quantum measurement.  The main challenge in probabilistic quantum metrology is to find the optimal generalized measurement that leads outcomes with super-precise estimates. The first key result is that  favourable outcomes cannot provide arbitrarily precise estimates, no matter how small their  probability is. This constitutes the \emph{ultimate quantum precision limit}. Of course, for a complete assessment,  the ultimate precision limit should be supplemented by the success probability (or, equivalently, by the abstention probability). In order to provide the most complete characterization of a probabilistic scheme that is amenable to optimization we follow a mini-max approach quantifying, for each possible value of the abstention probability, the worst precision among the favourable outcomes.

Finally, we want to stress that understanding the power of probabilistic operations in  general quantum tasks is a highly non-trivial and relevant problem in quantum information sciences. Probabilistic operations introduce a very particular non-linearity  (through normalization), which is in stark contrast with the linearity of any quantum deterministic operation.  Many no-go theorems stem from the linearity of quantum mechanics, and probabilistic operations might revoke them, turning the once-thought impossible into possible. For instance, it is well known that non-orthogonal states cannot be distinguished perfectly. A deterministic protocol that minimizes the average probability of error, produces two outcomes (one for every state). Each outcome has a given probability to give the incorrect answer, and at first sight it seems impossible to reduce  it beyond its optimal value. However, if a third outcome (abstention) is included, it is possible reduce the probability of an erroneous identification down to zero \cite{bagan_optimal_2012,fiurasek_optimal_2003}.  Similarly, the limitations imposed by the no-go theorem of realizing Bell measurements by linear-optical elements \cite{lutkenhaus_bell_1999} was overcome by the probabilistic KLM scheme \cite{knill_scheme_2001}. In quantum computing, most quantum algorithms require a certain (bounded) error probability. Exact algorithms that  provide a quantum speed-up and output the correct answer with certainty  are hard to come by \cite{ambainis_exact_2014,montanaro_exact_2013}.  Also, closely related to the current work (see discussion) we find  probabilistic amplification \cite{ferreyrol_implementation_2010, xiang_entanglement-enhanced_2011,zavattaa._high-fidelity_2011,kocsis_heralded_2013,chiribella_optimal_2013} and
weak-value amplification  \cite{dressel_colloquium:_2014,hosten_observation_2008,brunner_measuring_2010,zilberberg_charge_2011,jordan_technical_2014}.

In some of the above mentioned probabilistic schemes it is possible to devise means to pump up the success probability, while in quantum metrology it is often the case that in order attain the ultimate precision, the success probability is is doomed to be very small. Nevertheless, at a fundamental level, it is important to distinguish between ultimate versus de facto quantum limits. No matter how unlikely an event is, once it occurs it is a certainty; and certainties cannot violate ultimate bounds \cite{calsamiglia_quantum_2014}: faster than light signaling is impossible, winning the lottery (having bought the ticket) is not likely, but perfectly plausible. In the same way tunneling is highly improbable, but it can lead to spectacular or catastrophic consequences. This fundamental distinction has also motivated the definition of a complexity class in quantum computing \cite{aaronson_quantum_2005}. All in all, abstention can be considered as a resource per se in quantum information tasks, and this paper is devoted to the study of its power in metrology tasks in realistic noisy scenarios.

\section{Results}

\begin{figure}[htbp] 
   \centering
   \includegraphics[width=3.1in]{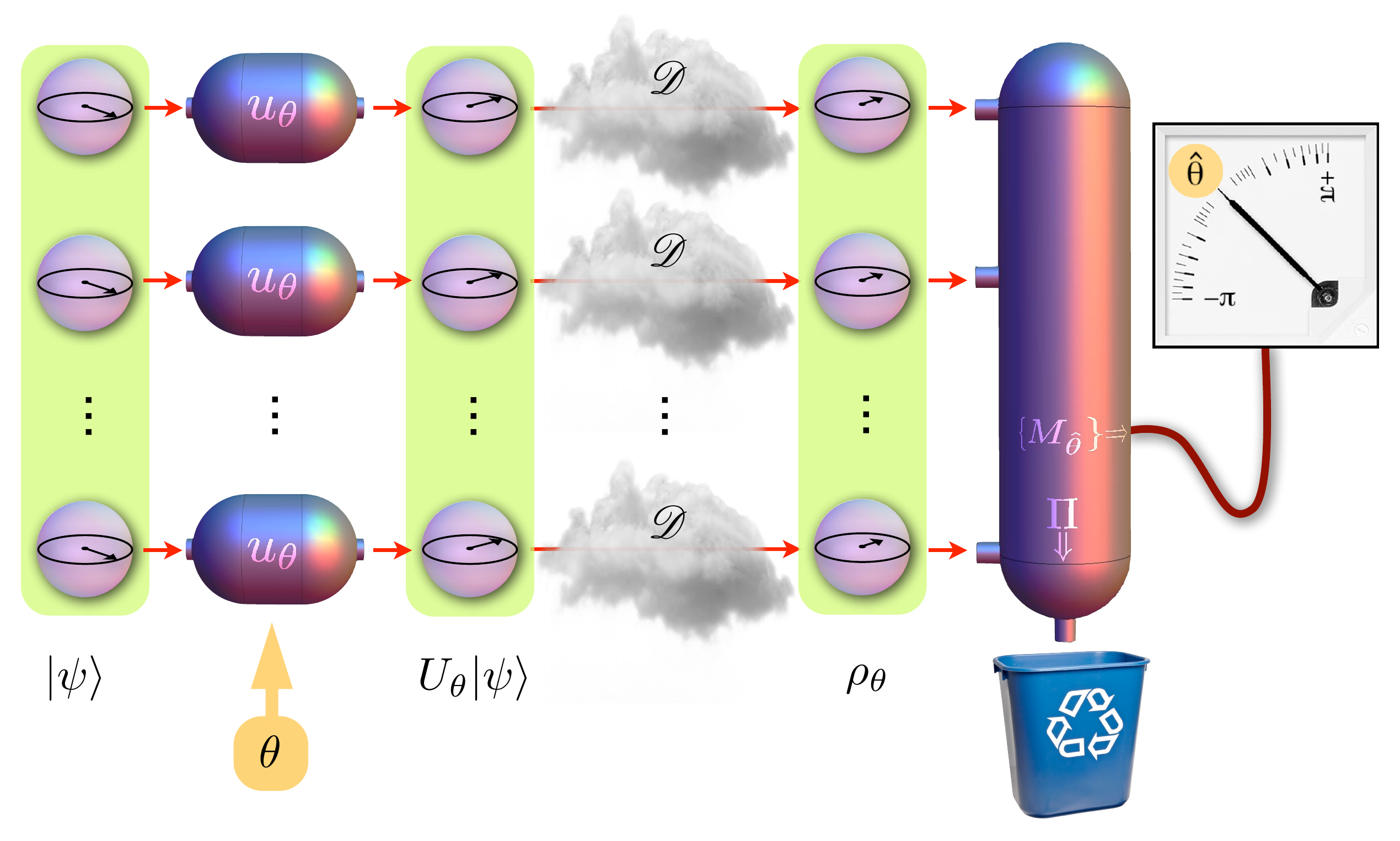} 
   \caption{{\bf Probabilistic Metrology protocol.} Pictorial representation of a probabilistic metrology protocol with $n$ qubits (depicted by small Bloch spheres). The probe state $|\psi\rangle$, which needs not necessarily be a product of identical copies, undergoes an evolution $U_\theta=u^{\otimes n}_\theta$ controlled by the unknown parameter $\theta$. Experimental noise $\mathscr D$ decoheres the system  before a collective measurement on \emph{all} qubits is performed. The measurement apparatus  either returns an ultra-precise estimate~$\hat{\theta}$ of the parameter or shows a failure signal. In the event of a failure, some information could be in principle scavenged (see last section in Results).}
   \label{fig:scheme}
\end{figure}

\noindent\textbf{\boldmath Optimal probabilistic measurement for $n$-qubits.} 
In the scope of this paper, metrology aims at estimating the parameter $\theta$ that determines the unitary evolution, $U_\theta:=u^{\otimes n}_\theta$, of a probe system of $n$ qubits in the presence of local decoherence, where $ u_\theta=\exp (i \theta\ketbrad{1}) $.

As depicted in Fig.~\ref{fig:scheme}, the initial $n$-partite pure state~$|\psi\rangle\langle\psi|=\psi$ (this shorthand notation will be used throughout the paper) is prepared and is let evolve. The state is affected by uncorrelated dephasing noise, which can be modeled by independent  phase-flip errors occurring  with probability $p_f=(1-r)/2$ for each qubit.
Its action on the $n$-qubits is described by a  map ${ \mathscr D}$ that commutes with the Hamiltonian, so that it could as well be understood as acting before or during the phase imprinting process. 

Next, the experimentalist performs a suitable measurement on $\rho_\theta= {\mathscr D}(U_\theta\psi U^\dagger_\theta)$ and, based on its outcome, decides whether to abstain or to produce an estimate~$\hat\theta$ for the unknown parameter~$\theta$. Note that this decision is based solely on the outcome of the measurement as,  naturally, the actual value of $\theta$ is unknown to the experimentalist. Our aim is to find the optimal protocol, e.g.,  the measurement that gives the most accurate estimates for a given probe state and for a given maximum probability of abstention.

We quantify the precision of the estimated phase~$\hat\theta$ by the fidelity ~\mbox{$f(\theta,\hat\theta)=[1+\cos(\theta-\hat\theta)]/2$}, and to assess the performance of the protocol we use the worst-case fidelity over the possible values of the unknown parameter.
\begin{equation}
\label{fidelity-def} 
F   =  \inf_{\theta\in (-\pi,\pi] }  \int  {\rm d} \hat \theta  \,    p(\hat \theta | \theta , \mathrm{succ})  f(\theta,\hat\theta),
\end{equation} 
where $p(\hat \theta|\theta ,\mathrm{succ})$ is the probability of estimating $\hat \theta$ after a successful event when the true value is $\theta$. The fidelity~$F$ and the probability of success~$S$  will fully characterize our probabilistic metrology strategies. For covariant families of states, such as our noisy probes $\{\rho_\theta\}_{\theta\in(-\pi,\pi]}$, the worst-case fidelity is entirely equivalent to the average fidelity (see Methods). 
To facilitate comparison with previous point-wise results, we present ours in terms of a scaled `infidelity', $\mse :=4(1-F)$, which approximates the mean-square error when the distribution 
$p(\hat\theta|\theta,\mathrm{succ})$  becomes peaked around the true value~$\theta$~\cite{berry_optimal_2012}.

Because of the symmetry of the problem, there is no loss of generality 
in choosing the covariant measurement defined by~$\{M_{\hat\theta}=U_{\hat\theta}\Omega U^\dagger_{\hat\theta}/(2\pi)\}_{\hat\theta\in(-\pi,\pi]}$, where~$\Omega$
is the so-called {\em seed} of the measurement. 
In addition, we have the invariant measurement operator \mbox{$\Pi=\openone- \int_0^{2\pi}d\hat\theta/(2\pi) U_{\hat\theta}\Omega\, U^\dagger_{\hat\theta}\le\openone$} that corresponds to the abstention event. 
With this, finding the optimal estimation scheme reduces to finding the operator $\Omega$  that maximizes the fidelity 
\begin{eqnarray}
F({S})&=&{1\over S}\max_{\Omega}\int{d\hat{\theta}\over2\pi} f(0,\hat\theta)\,{\rm tr}\!\left(U_{\hat\theta}\Omega U^\dagger_{\hat\theta}\rho\right)
\label{F_max}
\end{eqnarray}
for a fixed success probability
\begin{equation}
S=\int{d\hat \theta\over2\pi}{\rm tr}\!\left(U_{\hat\theta}\Omega U^\dagger_{\hat\theta}\rho\right) 
\label{succ}
\end{equation}
In deriving Eq.~(\ref{F_max}) we have used covariance  to get rid of the infimum in \eqref{fidelity-def},  thereby  formally fixing the value of $\theta$ to zero, and have defined $\rho=\mathscr D(\psi)$ accordingly.

\medskip

\noindent\textbf{Symmetric probes.} 
We now focus on probe states consisting $n$-qubits that are initially prepared in a permutation invariant state. This family includes most of the states considered in the literature, our case-study of multiple copies of equatorial-states, and also, as we will show below, the optimal probe-state for probabilistic metrology. The input state is given 
by,
\begin{equation}
\label{eq:symm}
\ket{\psi}=
\sum_{m=-J}^{J} c_{m}\ket{J,m},
\end{equation}
where $J=n/2$ is the maximum total spin angular momentum (hereafter spin for short) of~$n$ qubits 
and the set of states \mbox{$\{\ket{J,m}\}_{m=-J}^J$} spans the fully-symmetric subspace.   Given the permutation invariance of the noisy channel, the state $\rho=\mathscr{D}(\psi)$ inherits the symmetry of the probe, and can be conveniently written in a block diagonal form in the total spin bases~\cite{cirac_optimal_1999,bagan_optimal_2006} (see Methods),  
 \begin{equation}\label{eq:rhoj}
 \rho=\sum_{j} p_{j} \rho^{j}\otimes \frac{\id_{j}}{\nu_{j}}, 
 \end{equation}
where the state $\rho^{j}$ has unit trace, $p_j$ is the probability of $\rho$ having spin~$j$, and $\id_{j}$ stands for the identity in the $\nu_j$-dimensional multiplicity space of the irreducible representation of  spin $j$. The  sum over $j$ in \eqref{eq:rhoj} runs from
 $j_\mathrm{min}=0$ ($j_\mathrm{min}=1/2$) for  $n$ even (odd) to the maximum  spin~$J$.   Similarly, the measurement operators, can be taken to have the same symmetry and thus be of the form $\Omega=\sum_{j}\ketbrad{\chi_{j}}\otimes \id_{j}$, where $\ket{\chi_{j}}=\sum_{m}f^{j}_{m}\ket{j,m}$, $0 \leq f^{j}_{m}\leq 1$.
The maximum fidelity $F(S)$ for a fixed probability of success $S$ can hence be expressed in terms of the fidelity $F_j(s_j)$  in each irreducible block and its corresponding success probability $s_{j}$ (see Methods),
\begin{equation}
\label{eq:Fs}
F(S)= \max_{s_{j}}  \sum_{j} \frac{p_{j} s_{j}}{S} F_j(s_{j}),\quad S=\sum_{j}p_{j}s_{j}
\end{equation}
%
where
\begin{eqnarray}
&\displaystyle F_j(s_{j})=\frac{1}{2}\!\!\left(\!1\!+\!\frac{1}{s_{j}} \max_{0 \leq f^{j}_{m}\leq 1} \sum_{m}\! f^{j}_{m} \rho^{j}_{m,m+1}  f^{j}_{m+1}\!\! \right),&\nonumber \\[.5em]
&\displaystyle\mbox{ subject to  }s_{j}=\sum_{m }(f^{j}_{m})^{2} \rho^{j}_{m,m}.&
\label{eq:fidj}
\end{eqnarray}

This formulation of the problem allows for a natural interpretation of the probabilistic protocol as a two step process: i) a stochastic filtering channel
\begin{equation}
\label{eq:stoch}
{\mathscr F}(\rho)=\Phi\, \rho\, \Phi,\qquad \Phi=\sum_{j,m}(f^{j}_{m})^2\ketbrad{j,m}\otimes \id_{j},
\end{equation}
that  coherently  transforms each basis vector as \mbox{$\ket{j,m}\to f^{j}_{m}\ket{j,m}$}, so that it modulates the input to a state with enhanced phase-sensitivity, followed by ii)~a canonical covariant  measurement with seed \mbox{$\tilde\Omega=\sum_{j}\sum_{m,m'}\ket{j,m}\!\bra{j,m'}\otimes \id_{j}$} performed on the transformed state from which the value of the unknown phase is estimated.

By defining the vector $\xi^j$ with components given by $\xi^{j}_{m}=f_{m}^{j} (\rho^{j}_{m,m}/s_{j})^{1/2}$ and introducing the tridiagonal symmetric matrix $H^j$, with entries
\begin{eqnarray}
H^{j}_{m,m'}&=&2\delta_{m,m'}-a^{j}_{m}\delta_{m,m'-1}-a^j_{m'}\delta_{m-1,m'},\nonumber \\[.0em]
a^{j}_{m}&=&\frac{\rho^{j}_{m,m+1}}{\sqrt{\rho^{j}_{m,m}\rho^{j}_{m+1,m+1}}},
\label{eq:H^j_mm}
\end{eqnarray}
%
 we can easily recast the former optimization problem as,
\begin{eqnarray}
\label{eq:fidjmax}
& \kern-2em\displaystyle F_j(s_j)=1-\frac{1}{4}\msej,\quad  \msej :=\min_{\ket{\xi^{j}}} \bra{\xi^{j}}H^{j} \ket{\xi^{j}},  &\\[.5em]
 \label{eq:conds}
& \kern-2em\displaystyle \mbox{ subject to  } \braket{\xi^{j}} {\xi^{j}}\!=\!1 \mbox{ and } 0\!\le\! \xi^{j}_{m}\!\leq\! (\rho^{j}_{m,m}/\!s_{j})^{1/2}\!,& 
\end{eqnarray}
Note that $a^j_m$, and in turn $H^j$, depend on the strength of the noise but take the same values for {\em all} symmetric probe states, for we have that $\rho^{j}_{m,m'}\propto c_{m}c_{m'}$. 
%
%
For deterministic strategies ($S=1$, i.e., $s_{j}=1$ for all $j$) no minimization is required and one only needs to evaluate the expectation values of $H^{j}$ for the `state' $\xi^{j}_{m}=(\rho^{j}_{m,m})^{1/2}$.
For~large enough abstention, 
the problem becomes an unconstrained minimization, so  $\msej$  is the minimal eigenvalue of $H^{j}$, and $\ket{\xi^{j}}$  its corresponding eigenstate.   From \eqref{eq:conds} we find that the corresponding filtering operation only succeeds with a probability \eqref{eq:conds}
\begin{equation}\label{eq:Scrit}
S^{*}=\sum_{j}p_{j}s_{j}^{*} ,\qquad  s_{j}^{*} =\min_{m}\frac{\rho^{j}_{m,m}}{{\xi^{j}_{m}}^{2}} .
\end{equation}	
We will refer to $S^{*}$ as the critical success probability, since the fidelity will not improve by decreasing the success probability below this value, $F(S)=F(S^{*})$ for $S\leq S^{*}$.

\medskip

\noindent\textbf{Asymptotic scaling: particle in a potential box}. In order to compute the scaling of the precision as the number of resources becomes very large we need to solve the above optimization problem in the asymptotic limit of $n\to \infty$. We start be analysing the fidelity $F_j(s_{j})$ for blocks of large $j$.
As shown in Methods,  for each such block we define the ratios $x= m/j$, $m=-j,-j+1,\dots,j$, that approach a continuous variable  as~$j\to\infty$. In this limit, $\{\sqrt j \xi_m^{j}\}$ approaches a real function of $x$, $\sqrt j \xi_m^{j}\to\varphi(x)$, and the expectation value~\eqref{eq:fidjmax} becomes,
\begin{eqnarray}
\label{eq: sigmaj2}
 \msej &=&{
1\over j^2}\min_{\ket{\varphi}}%
\int_{-1}^1 \!\!dx  \left\{ \left[{d\varphi(x)\over dx}\right]^2\!\!+{V^j(x)}\varphi(x)^{2}\right\},\nonumber\\[.5em]
:\!&=&{
1\over j^2}\min_{\ket{\varphi}}\bra{\varphi} {\cal H}^j\ket{\varphi},
\end{eqnarray}
where we have dropped some boundary terms that are irrelevant for this discussion,  ${\cal H}^j:=-d^2/dx^2+V^j(x)$ plays the role of a `Hamiltonian', with a `potential'
\begin{equation}
V^j(x)= 2 j^2 (1-a^j_m)= j{1-r^2\over2r\sqrt{1-(1-r^2)x^2}} .
\label{eq:pot}
\end{equation}
Furthermore, in Eq.~(\ref{eq: sigmaj2}) the function $\varphi(x)$ must be also  differentiable and must satisfy the conditions
\begin{equation}
\label{eq:cond-cont}
\braket{\varphi} {\varphi}=\!\int_{-1}^1 \!\!\!dx\;[\varphi(x)]^{2}=1,\quad
\varphi(x)\leq \frac{\tilde\varphi(x)}{\sqrt{s_{j}}} ,
\end{equation}
%
where for a given large $j$ we define $\tilde\varphi(x)$  through 
\begin{equation}
\sqrt{j \rho^{j}_{mm}} \to \tilde\varphi(x),
\quad x=\frac{m}{j}
.\label{eq:kkt}
\end{equation}
%
It is now apparent from Eqs.~(\ref{eq: sigmaj2}) through~(\ref{eq:kkt}) that our optimization problem is formally equivalent to~that of finding the ground state wave-function of  a quantum particle in a box ($-1\le x\le1$) for a potential~$V^j(x)$ and subject to boundary conditions which are fixed by  the probe state,  the strength of the noise, and the success probability.  

\medskip

\noindent\textbf{Multiple-copies.} Although our Methods apply to general symmetric probes, for the sake of concreteness we study in full detail the paradigmatic case of a probe consisting of $n$ identical copies of equatorial qubits:
\begin{equation}
\label{eq:multicopy}
\ket{\psi_{{\rm cop}}}= \frac{1}{\sqrt{2^{n}}}( \ket{0}+\ket{1})^{\otimes n}. 
\end{equation}
%
Decoherence turns this symmetric pure state to a full rank state with a probability of having  spin $j$ given~by
 \begin{equation}\label{eq:pjmulticopy}
 p_{j}\simeq {{\rm e}^{-J{(j/J-r)^2\over 1-r^2}}\over\sqrt{\pi J(1-r^2)}},
 \end{equation}
which is valid around its peak at the typical value \mbox{$j_{0}=r J$}. 
For each irreducible block and before filtering we have a signal
\begin{equation}
\label{eq:gaussian}
\sqrt{j \rho^j_{mm}}\to\tilde\varphi(x)\simeq\left(\frac{j r}{\pi}\right)^{\frac{1}{4}}\expo{-\frac{ r j}{2} x^{2}}
\end{equation}
that peaks at $x=0$ with variance $\mean{x^{2}}=({2 r j})^{-1}$.

For deterministic protocols ($S=1$)  the constraints completely fix the solution, $\varphi(x)=\tilde\varphi(x)$, and the precision is obtained by computing the `mean energy' \mbox{$\msej=\mean{{\cal H}^j}_{\tilde\varphi}/j^2$}, Eq.~\eqref{eq: sigmaj2}. For large $j$ it is meaningful to use the harmonic approximation $V^j(x)\simeq V^j_0+\omega_{j}^{2} x^2$, where $V^j_0=j(1-r^2)/(2r)$ and $\omega_{j}^{2}=j(1-r^2)^2/(4r)$. The leading contribution to $\msej$ comes from the `kinetic' energy [i.e., the first term in~(\ref{eq: sigmaj2})]: $\mean{p^{2}}_{\tilde\varphi}=(1/4) \mean{x^{2}}^{-1}=j r/2$, whereas the harmonic term gives a sub-leading contribution. One easily obtains $\msej=(2 j r)^{-1}$. The leading contribution to the precision of the deterministic protocol is given by $\msej$ at the typical  spin $j_0$: $\sigma^2_{\rm det}=(2Jr^2)^{-1}=(nr^2)^{-1}$, in agreement with the previous known (point-wise) bounds (see Methods).

%

For unlimited abstention in a block of given spin~$j$ ($s_{j}$ very small) the minimization in~(\ref{eq: sigmaj2}) is effectively unconstrained and the solution 
(the filtered state) is given by the ground state $\varphi^{\rm g}(x)$ of the potential $V^j(x)$.
Within the harmonic approximation,  we notice that the effective frequency of the oscillator grows as $\sqrt{j}$, and the corresponding gaussian ground state is confined around~$x=0$ with variance $\mean{ x^{2}}=(r/j)^{1/2}(1-r^2)^{-1}$. In this situation both the kinetic and harmonic  contributions to the `energy' are sub-leading ---and so are the higher order corrections to $V^j(x)$. Thus, the precision $\msej$ for  a  spin~$j$ is ultimately limited by the constant term~$V^j_0$ of the potential. Up to sub-leading order one obtains 
  $\msej=(1-r^2)(2jr)^{-1}[1+(r/j)^{1/2}]$. The filtering of~$\tilde\varphi(x)$ to produce the  gaussian ground state $\varphi^{\rm g}(x)$ succeeds with probability  $s^{*}_{j}\sim \expo{-2 j \log(1+r)}$ (see Methods). Note that in the absence of noise ($r=0$) the potential $V^j(x)$ vanishes and the ground state is solely confined by the bounding box~$-1\le x\le1$. Then,  $\varphi^{\rm g}(x)=\cos(\pi x/2)$, which results in a Heisenberg limited precision (the ultimate pure-state bound)~$\mse=\pi^2/n^2$~\cite{summy_phase_1990,gendra_quantum_2013}.

 If the optimal filtering is performed on typical blocks,  $j \approx j_0$, one obtains $\mse=(1-r^{2})/(n r^{2})$, which  coincides with the ultimate deterministic bound found in~\cite{demkowicz-dobrzanski_elusive_2012,knysh_true_2014}.  
This shows that a probabilistic protocol performed on the uncorrelated multi-copy probe state $|\psi_{\rm cop}\rangle$ can attain the  
 precision bound of a deterministic protocol that requires a highly entangled probe. This bound is attained for a critical success probability \mbox{$S^{*}\simeq s^*_{j_{0}}\sim  \expo{-n r \log(1+r)}$}.
 
 More interestingly, we can push the limit further by post-selecting on the block with highest spin (by choosing~$f^{j}_{m}\propto \delta_{j,J}$) to obtain
 \begin{equation}
 \label{eq:ultimate}
\mse_{\mathrm{ult}}
:=\mse_{j\approx J}=\frac{1-r^{2}}{n r}  \left(1+\sqrt{\frac{2 r}{n}}\right),
  \end{equation}
  %
 with a critical probability given by $S^{*}= p_{J} s^{*}_{J}\sim\expo{-n  \log 2}$, independently of the noise strength.
 
Having understood the two extreme cases, i.e., the deterministic  ($S=1$) and unlimited-abstention protocols, we can quantify now the asymptotic scaling for an arbitrary success probability $\mse({S})$. The so-called complementary slackness condition~\cite{boyd_convex_2004,gendra_optimal_2013}, which  follows from the Karush-Kuhn-Tucker constrains [second inequality  in~\eqref{eq:cond-cont}] guarantees that, for a given value of $s_{j}$, 
the solution $\varphi(x)$ to~(\ref{eq: sigmaj2}) either saturates the above inequality ---in a region called coincidence set--- or it must take the value of an eigenfunction of the Hamiltonian  ${\cal H}^j$ defined after Eq.~(\ref{eq: sigmaj2}). The continuity of~$\varphi(x)$ and its derivative provide some matching conditions at the border of the coincidence set and a unique solution can be easily found. 

\begin{figure}[htbp] 
  \centering
   \includegraphics[width=3.4in]{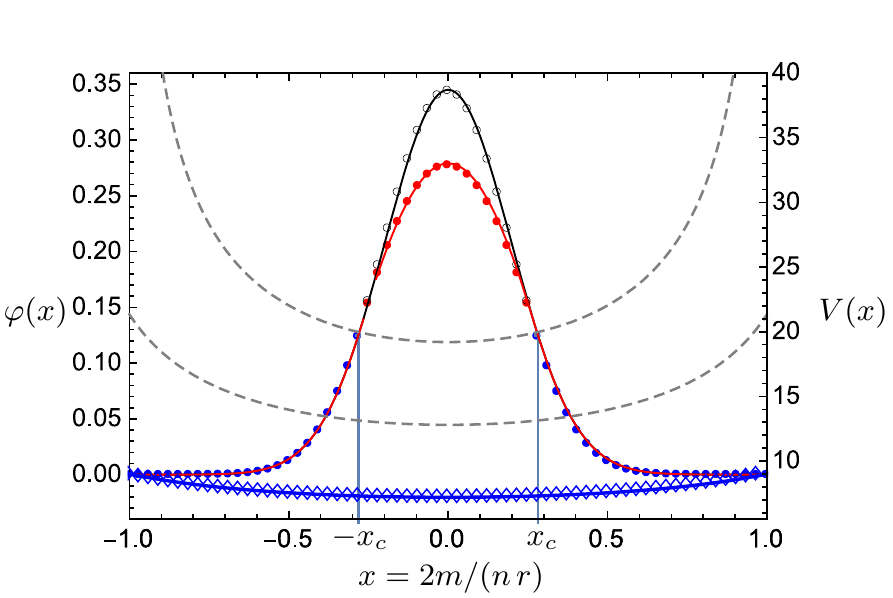} 
   \caption{{\bf Potential box equivalence}: Computing the action of the probabilistic filter and its precision is formally equivalent to 
   computing the the ground state and energy of a particle in a one-dimensional potential box. The  state $\tilde\varphi(x)$  (empty circles) before the probabilistic filter and the state $\varphi(x)$ (solid circles) after the filter are represented together with the  potential $V(x)$ (diamonds) corresponding to $j=n r/2$, see \eqref{eq:pot}, for success probability $S=0.75$, noise strength $r=0.8$ and $n=80$ probe copies. The unfiltered state (empty circles) has been rescaled so that it coincides with the filtered state in the region $|x| \geq x_{c} = 9/32$. The effective potential depends on the noise strength, as illustrated by the two additional dashed curves: for $r=0.2$ (above) and $r=0.6$ (below). Numerical (symbols) and analytical results (lines) are in full agreement.}
   \label{fig:profile}
\end{figure}

As shown in Figure 2, in the case of multiple copies the tails of $\varphi(x)$ coincide with the gaussian profile in~(\ref{eq:gaussian}) scaled by the factor $s_j^{-1/2}$ for $|x|>x_{c}$ (in the coincidence set), while the filter takes an active part in reshaping the peak into the optimal profile (for $|x|<x_{c}$). Clearly, the wider the filtered region, the higher the precision and the abstention rate. 
A simple expression for the leading order can be obtained if we notice that with a finite abstention probability one can change the variance of the wave function in~\eqref{eq:gaussian} but not its $1/j$ scaling. Hence, as for the deterministic case, only the kinetic energy and the constant term $V^j_0$ of the potential play a significant role. The solution can then be easily written in terms of the pure-state solution~\cite{gendra_optimal_2013}, which corresponds to a zero potential inside the box \mbox{$-1\le x\le1$}:
\begin{equation}
\label{eq:finiteQ}
\mse\simeq \mse_{j_{0}}=\frac{1-r^{2}}{n r^{2}}+r \mse_{\rm pure}(S)\approx\frac{1-(r^2/2) \bar S}{n r^{2}},
\end{equation}
where $\bar S:=1-S$ is the probability of abstention, $\mse_{\rm pure}$ is the precision for pure states ($r=1$) and for an effective number of qubits $n_{\rm eff}=2j_0$. The pre-factor~$r$ takes into account the scaling of the variance of the state~\eqref{eq:gaussian} as compared to the pure-state case.
The first equality of~\eqref{eq:finiteQ} uses the fact that for finite $S$, only abstention on blocks  about the typical spin~$j_{0}$ is affordable. This also fixes the value of $S$ to be approximately~$s_{j_{0}}$. The simple expression given in the last term in $\eqref{eq:finiteQ}$ is not an exact bound, but does provide a good approximation for moderate values of $\bar S$ (see Figure 3).
We notice that for low levels of noise ($r\approx 1$) one can have a considerable gain in precision already  for finite abstention.
\begin{figure}[htbp] 
   \centering
   \includegraphics[width=3.4in]{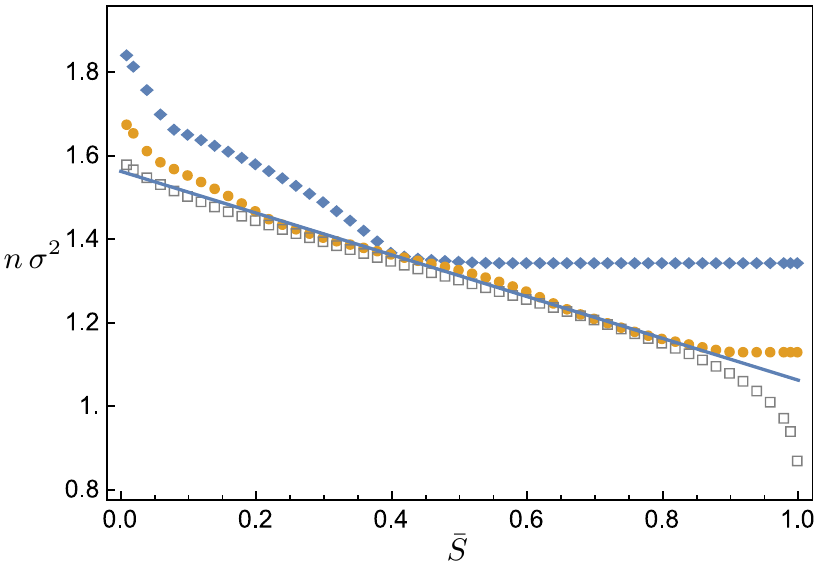} 
   \caption{{\bf Small \boldmath{$n$} precision.} 
   Numerical results for the rescaled precision $n \sigma^2$ for a noise strength of $r=0.8$ as a function of the abstention probability $\bar S=1-S$ for various numbers of copies $n =\{6,10,20\} $ (diamonds, circles and squares).  The critical success probability is clearly identified for the first ($n=6$) curves at $\bar S^*=0.46$. The solid line is the approximated analytical result \eqref{eq:finiteQ}.}
   \label{fig:sigmaVSQ}
\end{figure}

 \noindent\textbf{Finite \boldmath{$n$}.} Up to this point, we have given analytical results for asymptotically large~$n$, the number of resources. In order to get exact values for finite $n$ we need to resort on numerical analysis. The main observation here is that  our optimization problem  can be cast as a semidefinite program: 
 \begin{equation}
 \label{eq:SDP}
 \mse=\min_{\Lambda\,:\,{\cal C}} \tr{H\Lambda}
 \end{equation}
 subject to a set of linear conditions on the matrix~$\Lambda$: \mbox{${\cal C}:=\{ \Lambda\geq 0,\, {\rm tr}\,\Lambda=1,\, \Lambda^{j}_{mm}\leq p_{j} \rho^{j}_{m,m}/S\}$}, where $\Lambda$, as well as $H$, have the block diagonal form: $\Lambda=\oplus \Lambda^{j}$ and $H=\oplus H^{j}$. Semidefinite programming problems, such as this, can be solved efficiently and with arbitrary precision~\cite{boyd_convex_2004}.

Figure 3 shows representative results for moderate  (experimentally relevant) number $n$ of qubits for the precision as a function of the abstention probability and noise strength $r=0.8$. We observe that for small values of $n$ the precision decreases quite rapidly until the critical~$S^{*}$ after which the precision cannot be improved. For larger~$n$ the initial gain is less dramatic, but the critical  point (or plateau) is reached for higher abstention probabilities, hence allowing to reach lower precision rates. We see that for moderately large $n$, abstention can easily provide $60\%$ improvement of the precision.

Figure 4 shows the scaling of the precision with the amount of resources ($n$)  for low levels of noise $r=95\%$ and for different values of the abstention probability $\bar S$. For low~$n$ all curves exhibit a similar (SQL $n^{-1}$  scaling). Very soon  the curve corresponding to unlimited abstention shows a big drop with a quantum-enhanced transient scaling: $n^{-(\alpha+1)}$, where $\alpha>0$ depends on the noise strength. This curve saturates for very large $n\sim 500$ to the ultimate asymptotic limit \eqref{eq:ultimate}, which is has again SQL scaling. The curves for finite $S$ follow closely the optimal scaling up to the point where they meet the asymptotic curve \eqref{eq:finiteQ}.
The larger the abstention probability, the later this transition happens. In addition, in the figure the ultimate scaling for $r=99\%$ is shown to illustrate that for weaker noise levels the transient is more abrupt (larger $\alpha $).

\begin{figure}[htbp] 
   \centering
   \includegraphics[width=3.4in]{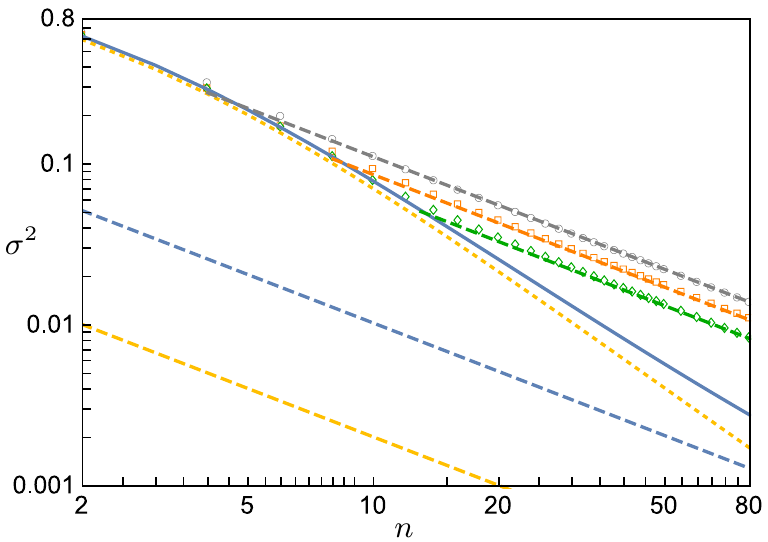} 
   \caption{{\bf \boldmath Moderate and large $n$ scaling.}  Ultimate precision scalings (for $n\to \infty$) are of fundamental interest, however  from practical perspective understanding the transient behaviour equally important. The plot shows the precision $\mse$ for different abstention probabilities  $\bar S=\{0,0.5,0.9\}$   (from top to bottom) for $r=0.95$. The fourth line (blue) shows the exact ultimate limit ($\bar S$ arbitrarily small). The dashed lines show the corresponding asymptotic limits given in Eqs. \eqref{eq:finiteQ} and \eqref{eq:ultimate}.  In addition, for weaker noise strength $r=0.99$
    we show lines corresponding to the ultimate precision (in yellow).}
   \label{fig:sigmaVSQ}
\end{figure}

\medskip

\noindent\textbf{Ultimate bound for metrology}.
So far we have studied the best precision bounds that can be attained for a fixed input state.
A very relevant question of fundamental and practical interest is whether this bounds can be overcome by an appropriate choice of  such state. We answer this question in the negative: the precision bound in Eq.~\eqref{eq:ultimate} is indeed the ultimate bound for metrology~in the presence of local decoherence and can only be attained by a probabilistic strategy.

To this aim, we first show in Methods that for any probe state and any measurement that attain a fidelity $F$ with success probability $S$, we can find  a new probe lying in the fully symmetric subspace ($j=J$) and a permutation invariant measurement that attain the very same fidelity with the very same success probability. This shows that the formulation that we have introduced,  which deals with such probes, is actually completely general. 

We now recall that the Hamiltonian  ${\cal H}^j$ is independent of the choice of probe state and that such choice determines only the shape of the state~$\tilde\varphi(x)$ before filtering, and the probability $p_{j}$ of belonging to the subspace of  spin $j$. Since the
bound~\eqref{eq:ultimate} is attained by the ground-state $\varphi^{\rm g}(x)$ of the potential $V^J(x)$, the choice of probe cannot further improve the precision, but only change the success probability.
In particular one might increase $S$ by choosing a probe state that gives rise to a profile $\tilde\varphi(x)=\varphi^{g}(x)$ for $j=J$, without any  filtering within the block. In this case the critical success probability becomes $S^*=p_{J}={\rm e}^{-n[\log 2-\log(1+r)]}$ (see Methods), which is larger than that attained by $|\psi_{\rm cop}\rangle$. 

At the other extreme, for deterministic strategies, the calculation of $\sigma^2_{\rm opt}(1)$ can be easily carried out performing first the sum over $j$ and then optimizing over the $(n+1)$-dimensional probe state.  In the continuum limit (for large $n$) such calculation  can again be cast as  a variational problem formally equivalent to that of finding the ground-state of particle in a box with the harmonic potential $V(y)= n r^{-2}({1-r^{2}})(1+ y^{2})$, $-1\le y=m/J\le 1$. 
The corresponding ground state wave function and its energy provide the optimal probe state and precision respectively:
\begin{equation}
\label{eq:probe}
\psi_{\mathrm{op}}(y)=\left[{n(1-r^2)\over(2\pi r)^{2}}\right]^{\frac{1}{8}}\expo{-{\sqrt{n(1-r^2)}\over4r}y^2} 
\end{equation}
and
\begin{equation}
\mse_{\mathrm{op}}(1)={1-r^2\over n r^2}+{2\sqrt{1-r^2}\over n^{3/2} r}.
\label{eq:opten}
\end{equation}
These results agree with their pointwise counterparts in~\cite{demkowicz-dobrzanski_elusive_2012,knysh_true_2014}. Quite surprisingly the presence of noise brings the pointwise and global approaches in agreement, as to both the attainable precision and the optimal probe state are concerned.
This is in stark contrast with the noise-less case where the probe  $\psi(y)=\cos(y\pi/2)$ is optimal for the global approach and gives $\mse_{\rm opt}=\pi^{2}/n^{2}$, while
the NOON-type state $\ket{\psi}=2^{-1/2}(\ket{J,J}+\ket{J,-J})$ provides the optimal point-wise precision $\mse_{\rm opt}=1/n^{2}$.

It remains an open question to find the optimal probe state given a finite values of $\bar S$.  As argued above, a finite $S$ will only be able to moderately reshape the profile without significantly changing the scaling of its width. Therefore we expect the optimal state to be fairly independent of the precise (finite) value of $S$, and hence very close to that obtained for the deterministic case ($S=1$). Numerical evidence (optimizing simultaneously over probes and measurements) suggests that this is indeed the case provided $S$ is not too small. With this we are lead to conjecture that the optimal probe state is given by 
\begin{equation}
\label{eq:optFiniten}
c^{\mathrm{opt}}_{m}\propto\cos\left(\frac{m \pi}{n+2} \right) \expo{-{\sqrt{\frac{1-r^{2}}{ r^{2 } n^{3}}}} m^2}
\end{equation}
independently of $S$  (finite),  which  agrees with \eqref{eq:probe}  for asymptotically large $n$. Note that the cosine prefactor guarantees that the solution converges to the optimal one for $r\to 1$ and  it keeps the state confined in the box for all values of $n$ and $r$. Such states continue to have a dominant typical value of $j=j_{0}$ and in those blocks both the kinetic and harmonic contributions to the energy are  of sub-leading order. Hence,  for  probes of the form \eqref{eq:optFiniten} the enhancement due to abstention is very limited, up until very high abstention probabilities where one can afford to post-select high spin states to reach the ultimate limit \eqref{eq:ultimate}.

\medskip

\noindent\textbf{Scavenging information from discarded events}
The aim of probabilistic metrology is twofold. First, it should estimate an unknown phase $\theta$ encoded in a quantum state with a precision that exceeds the bounds of the deterministic protocols. Second,  it should assess the risk of failing to provide an estimate at all (e.g., it should provide the probability of success/abstention). Probabilistic metrology protocols are hence characterized by a precision versus probability of success trade-off curve~$F(S)$ or, likewise $\mse(S)$. As such, no attention is payed to the information on $\theta$ that might be available after an unfavorable outcome.  
Here, we wish to point out that one can attain $\mse_{\rm opt}(S)$ and still be able to recover, or scavenge, a fairly good estimate from the discarded outcomes (see Fig.~\ref{fig:scheme}). 

The optimal scavenging protocol can be easily characterized in terms of the stochastic map~${\mathscr F}$ in~\eqref{eq:stoch}, which describes the state transformation after a favorable event, and that associated to the unfavorable events: 
\begin{equation}
{\bar {\mathscr F}}(\rho_{\theta})=\bar\Phi \rho_{\theta} \bar\Phi ,\quad \bar\Phi=\sum_{j,m}(\bar{f}^{j}_m)^2\ketbrad{j,m}\otimes \id_{j} ,
\end{equation}
where the weights $\bar{f}^{j}_{m}$ are defined through the equation~$(\bar{f}^{j}_{m})^2=1-(f^{j}_{m})^2$. The addition of the two stochastic channels, $\bar{\mathscr F}+{\mathscr F}$, is trace-preserving, i.e., it describes a deterministic operation, with no post-selection. The final measurement is given by the seed~$\tilde\Omega$ defined after~(\ref{eq:stoch}) for both favorable and unfavorable events. Thus, we can easily compute the precision ${\bar\sigma}^{2}(S)$  for the the latter, as well as the precision~$\sigma_{{\rm all}}^{2}(S)$ when all outcomes are considered. Clearly, we must have that $\sigma_{{\rm all}}^{2}(S)\geq \mse_{\rm det}$~\cite{combes_quantum_2014}, as $\mse_{\rm det}$ refers to the optimal deterministic protocol.
\begin{figure}[htbp] 
   \centering
   \includegraphics[width=3.4in]{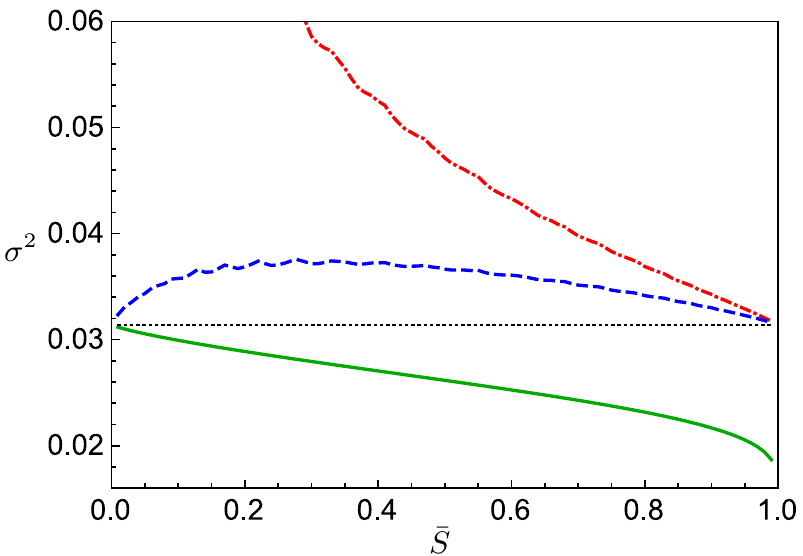} 
   \caption{{\bf Scavenging information.} Precision $\sigma^2$ vs probability of abstention $\bar S:=1-S$ from numerical optimization for $n = 50$ and $r = 0.8$.  The green solid (red dash-dotted) correspond to $\sigma^2_{\rm opt}$ ($\bar\sigma^2$), where only the favorable (unfavorable) events are taken into account. The dashed curve~$\sigma_{\rm all}^2$, where an estimate is provided on \emph{all} outcomes, favorable or unfavorable. For low success probability ($\bar S$ close to unity), both~$\bar\sigma^2$, and~$\sigma_{\rm all}^2$, approach the precision of the deterministic protocol~$\sigma^2_{\rm det}$ (dotted line).}
   \label{fig:scav}
\end{figure}

As shown in Figure~\ref{fig:scav} a protocol that is optimized with some probability of abstention $\bar{S}$, performs slightly worse when forced to provide always a conclusive outcome. In particular we notice that if such protocol is designed to work at the ultimate limit regime, with precision~$\mse_{\mathrm{ult}}$, which requires a very large abstention probability ($S\to 0$)~\cite{combes_quantum_2014}, its  performance coincides that of the optimal deterministic protocol.  Actually, this observation follows (see Methods) from Winter's gentle measurement lemma (Lemma~9 in~\cite{winter_coding_1999}), that states that a measurement with a highly unlikely outcome causes only a little disturbance to the measured quantum state. This is in contrast to the claims in~\cite{combes_quantum_2014}, where a random estimate is assigned to the discarded events.

\section{Discussion}

We have shown that abstention or post-selection can  counterbalance the adverse errors in a noisy metrology task. Our results are theoretical and concern abstract systems of $n$  qubits. However, they apply  to different quantum metrology implementations, ranging from  Ramsey interferometry for frequency standards \cite{bollinger_optimal_1996,huelga_improvement_1997}, atomic magnetometry \cite{budker_optical_2007,napolitano_interaction-based_2011}, and quantum photonics (single or multi-mode setups) where the number operator introduced here will play the role of number of photons.

Post-selection is already widely used for preparing quantum information resources, e.g. single photons from weak coherent pulses, heralded down-conversion for EPR-type states, or NOON states for metrology applications. Although some degree of post-selection is common in experiments, its tailored optimised use is not fully exploited. Only recently there have been important developments  
in this direction in the context of weak value amplification \cite{dressel_colloquium:_2014,ferrie_weak_2014,hosten_observation_2008} -- we note on passing that these schemes can be considered a particular instance of our general set-up, and hence are subject to our bounds.

The optimal probabilistic measurement presented here can be understood as a filtering process selecting the total angular momentum followed by a modulating filter, and a final standard covariant phase measurement. 
The latter can be implemented by the (almost) optimal adaptive scheme proposed in \cite{berry_optimal_2000}. The modulation could be implemented by sequential use of amplitude-damping channels taking inspiration from recent experiments in state amplification \cite{ferreyrol_implementation_2010,zavattaa._high-fidelity_2011,kocsis_heralded_2013}. In implementations that allow for an individual control of the qubits, such as ion traps, the projection onto the angular momentum basis can be efficiently carried out \cite{bacon_efficient_2006}. 
For implementations with less degree of control, the projection onto the fully symmetric sub-space can, as a last resort, be implemented by post-selecting outcomes with this symmetry. For instance a simple Stern Gerlach measurement could lead to outcomes ($m=J$) with a precision beyond the deterministic limits.

Regarding the implementation of our conjectured optimal probe state one can use available non-linear $N^{2}$-type two-body interactions to turn the multi-copy gaussian profile to the wider optimal gaussian. Although, our case-study focuses on local dephasing noise, 
our Methods can be adapted and similar, if not greater, benefits are expected for more general and implementation-specific noise models including correlated noise.

In conclusion, we have shown what are the ultimate limits in precision reachable by any (deterministic or stochastic) quantum metrology  protocol in a realistic scenario with local decoherence. We have derived the optimal bounds that can be reached when a certain rate of abstention is allowed and hence provided a full assessment of the risks and benefits of the probabilistic strategy.
The benefits are clear for finite and for  asymptotically large number of copies, and the  precision is strictly better than that attained by deterministic strategies, which include optimal preparation of probe states. The ultimate quantum metrology scaling limit is only reached with a large abstention rate. However, in that case we have shown that it is possible to obtain estimates with  standard (deterministic) precision from the discarded events. In this sense, seeking ultra-sensitive measurements is a low-risk endeavour.

\section{Methods}

\noindent{\bf Notation.}
Throughout this section we use the following notation. The $n$-qubit computational basis is denoted by $\{| b\rangle\}_{b=0}^{2^n-1}$, where~$b=b_1b_2\cdots b_n$ is a binary sequence, i.e., $b_i=0,1$ for $i=1,2,\dots,n$. We denote  by $|b|$ the sum of the $n$ digits of $b$, i.e.,  $|b|:=\sum_{j=1}^n b_j$. The digit-wise sum of $b$ and $b'$ modulo~$2$ will be simply denoted by $b+b'$, hence~$|b+b'|$ can be understood  as the Hamming distance between $b$ and $b'$, both viewed as binary vectors.

The permutations of $n$ objects, i.e., the elements of the symmetric group $S_n$, are denoted by $\pi$. We define the action of a permutation $\pi$ over a binary list~$b$ as \mbox{$\pi(b):=b_{\pi(1)}b_{\pi(2)}\cdots b_{\pi(n)}$}. This induces a unitary representation of the symmetric group on the Hilbert space~${\cal H}^{\otimes n}$ of the $n$ qubits through the definition \mbox{$U_\pi|b\rangle:=|\pi(b)\rangle$}. The (fully) symmetric subspace of~${\cal H}^{\otimes n}$, which we denote by ${\cal H}^{\otimes n}_+$, plays an important role below. An orthornormal basis can be labelled $\beta=|b|$, where $\beta=0,1,\dots,n$:
\begin{equation}
|\beta\rangle
=\begin{pmatrix}n\\\beta\end{pmatrix}^{-1/2}\sum_{b\in B_\beta}|b\rangle  \mbox{ with } \beta=0,1,\ldots n
\label{basis H_+}
\end{equation}
where $B_\beta=\{b : |b|=\beta\}$. It is well-known that the symmetric subspace~${\cal H}^{\otimes n}_+$ carries the irreducible representation of spin $j=J:=n/2$ of $SU(2)$. In this language, the magnetic number $m$ is related to~$\beta$ by $m=n/2-\beta$ (here we are mapping $\beta_i\to m_i=(-1)^{\beta_i}/2$ for qubit $i$). In other words, we map $|\beta\rangle\to|n/2,n/2-\beta\rangle$, where we stick to the standard notation $|j,m\rangle$ for the spin angular momentum eigenstates. 

We will be concerned with evolution under unitary transformations $U_\theta:=u^{\otimes n}_\theta$, where $u_\theta=\exp(i\theta |1\rangle\langle1|)$, $\theta\in(-\pi,\pi]$. The operator $N$ such that $U_\theta={\rm e}^{i\theta N}$  will be referred to as number operator for obvious reasons: $N|b\rangle=|b||b\rangle$. The effect of noise is taken care of by a CP map ${\mathscr D}$, so the actual evolution of an initial $n$-qubit state~$\psi:=|\psi\rangle\langle\psi|$ is $\psi\to {\mathscr D}(U_\theta\psi U^\dagger_\theta)=\rho_\theta$. 

With this notation the fidelity and success probability in Eqs.~(\ref{F_max}) and~(\ref{succ}) can be written as
\begin{eqnarray}
F({S})
&=&\frac{1}{2}\left(1+\frac{1}{S} \max_{\Omega} \sum_{b,b'} \Omega_{b,b'} \rho_{b',b}\delta_{|b'|,|b|+1}\right),
\label{F(S) b}
\\
S&=&
 \sum_{b,b'} \Omega_{b,b'} \rho_{b',b}\delta_{|b'|,|b|},
 \label{S b}
\end{eqnarray}
where the Kronecker delta tensors result from the integration of~$\hat\theta$. 
\medskip

\noindent{\bf Local dephasing: Hadamard channel.}
In this paper we consider uncorrelated dephasing noise, which can be modeled by phase-flip errors that occur with probability $p_f$. i.e., at the single qubit level, the effect of the noise is~$\varrho\to (1-p_f)\varrho+p_f\,\sigma_z\varrho\,\sigma_z$, where $\sigma_z$ is the standard Pauli matrix $\sigma_z={\rm diag}(1,-1)$. 
For states of~$n$ qubits, this, so called dephasing channel ${\mathscr D}$, is most easily characterized through its action on the operator basis~$\{|b\rangle\langle b'|\}_{b,b'=1}^{2^n-1}$ as 
\begin{equation}
{\mathscr D}(|b\rangle\langle b'|)=r^{|b+b'|}|b\rangle\langle b'|,
\end{equation}
where the parameter $r$  is related to the error probability~$p_f$ through $r=1-2p_f$. The effect of ${\mathscr D}$ on a general $n$-qubit state $\varrho=\sum_{b,b'}\varrho_{b,b'}|b\rangle\langle b'|$ can then be written as the Hadamard (or entrywise) product 
\begin{equation}
{\mathscr D}(\varrho)=\sum_{b,b'}r^{|b+b'|}\varrho_{b,b'}|b\rangle\langle b'|:={\cal D}\circ\varrho,
\label{Hadamard}
\end{equation}
where ${\cal D}:=\sum_{b,b'}r^{|b+b'|}|b\rangle\langle b'|$ and hereafter we understand that the sums over sequences run over all possible values of $b$ (and $b'$) unless otherwise specified. Note that Hadamard product is basis-dependent.

\medskip

\noindent{\bf Symmetric probes.} If the probe state is fully symmetric, i.e., $|\psi\rangle\in{\cal H}^{\otimes n}_+$, it can be written as $|\psi\rangle=\sum_{\beta} \psi_\beta |\beta\rangle$, where~$|\beta\rangle$  is defined in~(\ref{basis H_+}) and the components are related to those in~(\ref{eq:symm}) by $c_m=\psi_{J-m}$ and can be taken to be positive with no loss of generality (any phase can be absorbed in the measurement operators). Then,~$\rho={\mathscr D}(\psi)={\cal D}\circ\psi$ in~(\ref{F(S) b}) and~(\ref{S b}) becomes
\begin{equation}
\rho=\sum_{\beta,\beta'}{\psi_{\beta'} \psi_{\beta}\over {n\choose \beta}^{1/2}{n\choose \beta'}^{1/2} }\sum_{b\in B_\beta} \sum_{b'\in B_{\beta'}}\! r^{|b+b'|}|b'\rangle\langle b|  .
\end{equation}
Since $\rho$ is permutation invariant, $\Omega$ can be chosen to be so and we can easily write~(\ref{F(S) b}) and~(\ref{S b}) in the  spin basis. We just need the non-zero Clebsch-Gordan matrix elements~$\langle j,m'|b'\rangle\langle b|j,m\rangle$, where implicitly $m=J-\beta$, $m'=J-\beta'$. If we introduce the shorthand notation ${\cal D}^j_{m',m}:=\langle j,m'|{\cal D}|j,m\rangle$, then using~\cite{calsamiglia_local_2010}, we have
\begin{eqnarray}
{\cal D}^j_{m',m}&=&
\sum_{b\in B_\beta} \sum_{b'\in B_{\beta'}}\! r^{|b+b'|}\langle j,m'|b'\rangle\langle b|j,m\rangle\nonumber \\
&=&(1-r^2)^{J-j}r^{m-m'} \sum_{k}[\Delta_k^{(j)}]_m^{m'} r^{2k} ,
\label{eq:CG}
\end{eqnarray}
where
\begin{equation}
\big[\Delta^{(j)}_k\big]^{m'}_m\!\!:=\!{\sqrt{(j\!-\!m)!(j\!+\!m)!(j\!-\!m')!(j\!+\!m')!}\over
(j\!-\!m\!-\!k)!(j\!+\!m'\!-\!k)!(m\!-\!m'\!+\!k)!k!
} ,
\label{coeff Wigner}
\end{equation}
and the sums run over all integer values for which the factorials make sense. Recalling \eqref{eq:rhoj}, a simple expression, involving just a sum over $k$ in \eqref{coeff Wigner}, for \mbox{$\rho^j_{m,m'}=p_j^{-1} \mathrm{tr}(\ketbra{j,m}{j,m'}\otimes \id_j \; \rho)$} follows by combining the above results. In short,
\begin{equation}\label{eq:rhojmm}
\rho^j_{m',m}={1\over p_j}{c_{m'}c_m\over{n\choose J-m'}^{1/2}{n\choose J-m}^{1/2}} {\cal D}^j_{m',m},
\end{equation}
where
\begin{equation}\label{eq:pj}
p_j=\nu_{j}\sum_{m}{c_{m}^{2}\over{n\choose J-m}} {\cal D}^j_{m,m},
\end{equation}
and the multiplicity is given by,
\begin{equation}
\nu_j={n\choose J-j}{2j+1\over J+j+1}
\end{equation}
and $a^j_m$ in~(\ref{eq:H^j_mm}) becomes
\begin{equation}
a^j_m={{\cal D}^j_{m,m+1}\over\sqrt{{\cal D}^j_{m,m}{\cal D}^j_{m+1,m+1}}}\,.
\label{a^j_m D}
\end{equation}

\noindent{\bf{Relevant expressions for the multi-copy state.}} If the input state is of the form given in Eq.~\eqref{eq:multicopy}, the above expressions \eqref{eq:rhojmm} and \eqref{eq:pj} become

\begin{equation}
\label{eq:rhojmm2}
\rho^{j}_{m',m}={{\cal D}_{m',m}^j\over \sum_{m}{\cal D}_{m,m}^j}  \mbox{ and } p_j=\nu_j\;2^{-n}\sum_{m}{\cal D}_{m,m}^j\,,
\end{equation}
where
\begin{equation}
\label{eq:sumD}
\sum_{m}{\cal D}_{m,m}^j=(1-r^2)^{J-j}{(1+r)^{2j+1}-(1-r)^{2j+1}\over 2\;r}
\end{equation}

The probability to find the state in the fully symmetric subspace ($j=J$) is important when assessing the success probability of the the ultimate bounds.  Since the multiplicity for the maximum  spin $J$ is equal to one, it can be readily seen that $p_{J}$ scales as
\begin{equation}
p_J\sim {\rm e}^{-n[\log 2-\log(1+r)]}\,,
\end{equation}

The critical probability $s_{j}^{*}$ within a block can also be computed in the asymptotic limit $j\gg 1$ from  Eq.~\eqref{eq:Scrit} 
\begin{equation}
s_j^*={\rho_{j,j}^j\over (\xi_j^j)^2}\sim  \expo{-2j\log{1+r}}
\end{equation}
where $\xi_m^j$ is the gaussian ground state, with $(\xi_j^j)^2\sim \exp(-(1-r^2)\sqrt{j/4r})$. For $m=m'=j$ equation \eqref{eq:CG} gives $D^j_{j,j}=(1-r^2)^{J-j}$ which together with  \eqref{eq:rhojmm2} and \eqref{eq:sumD} gives  $\rho_{j,j}^j\sim \exp[ 2 j \log(r+1) ]$. This scaling dominates over that of $\xi_m^j$ , and hence determines the scaling of $s_{j}^{*}$.
From here we obtain critical value for the overall success probability $S^*=p_{J}\;s_{J}^*\sim\expo{-n\log\;2}$.

\medskip
\noindent{\bf{Ultimate bound without in-block filtering.}}
In the results section we discuss the possibility to prepare a probe state such that after the action of  noise becomes an optimal state within  the fully symmetric subspace $j=J$. Here we give what its critical success probability, which only entails computing $p_{J}$.

For this purpose we first recall that  the optimal filtered state $\xi^j_m$,  defined  before Eq.~(\ref{eq:H^j_mm}),  has to fulfil 
\begin{equation}
\xi^j_m=f^j_m c_m \sqrt{{\nu_j {\cal D}^j_{m,m}\over s_j p_j {n\choose J-m}}}\  .
\label{eq:xi expl}
\end{equation}
The probability of falling in the block of maximum  spin $J$ for a given filtered state~$\xi^j$ can be easily derived from~(\ref{eq:xi expl}) recalling that the probe state $\psi$ is normalized, and thus $\sum_{m}c_m^2=1$. Solving~(\ref{eq:xi expl}) for $c^2_m/p_J$ and summing over $m$ we obtain
\begin{equation}
{1\over p_J}=s_J\sum_{m=-J}^{J}{{n\choose J-m}\over{\cal D}^{J}_{m,m}}\left({\xi^J_m\over f^J_m}\right)^2.
\end{equation}
Now, for our strategy  all $j$ but the maximum one, \mbox{$j=J$}, are filtered out, and no further filtering is required within the block $J$, i.e. we have $f^J_m=1$, for all $2J+1$ values of $m$. Then $s_J=1$ and
\begin{equation}
p_J=\left\{\sum_{m=-J}^J {{n\choose J-m}\over{\cal D}^{J}_{m,m}}\left({\xi^J_m}\right)^2\right\}^{-1}.
\end{equation}
In the asymptotic limit the probability $p_J$ can be estimated by noticing that the optimal distribution $(\xi^j_m)^2$ is much wider than ${n\choose J-m}/{\cal D}^{J}_{m,m}$ and can be replaced by~$(\xi^J_0)^2$. Around $m=0$, we can use the asymptotic formulas
\begin{eqnarray}
{\cal D}^{j}_{m,m}&\sim& (1-r^2)^{J-j}{(1+r)^{2j+1}\over 2\sqrt{\pi r j}}{\rm e}^{-r m^2/j},\label{D^j_mm Gauss}\\[.5em]
{n\choose J-m}&\sim&{2^n\over\sqrt{\pi J}}{\rm e}^{-m^2/J}.
\end{eqnarray}
They can be derived using the Stirling approximation and saddle point techniques. Eq.~(\ref{D^j_mm Gauss}) also requires the Euler–Maclaurin approximation to turn the sum over $k$ in~(\ref{eq:CG})  into an integral that can be evaluated using again the saddle point approximation. Retaining only exponential terms, $S^{*}=p_J\sim(1+r)^n/2^n={\rm e}^{-n[\log 2-\log(1+r)]}$. 
\medskip

\noindent{\bf Performance metrics: Equivalence of worst-case and average fidelity, and point-wise vs. global approach} 
Here we give a simple proof that for the estimation problem at hand, the worst-case fidelity in Eq.~(\ref{fidelity-def}) and the average fidelity 
\begin{equation}
F_{\rm av}:=\int {d\theta\over2\pi} \int d\hat\theta \,p(\hat\theta|\theta,{\rm succ}) f(\theta,\hat\theta)
\label{Fav}
\end{equation}
(the integration limits $-\pi$, $\pi$ are understood)
take the same value, and so do the corresponding success probabilities. We recall that $f(\theta,\hat\theta)=[1+\cos(\theta-\hat\theta)]/2$ and $p(\hat\theta|\theta,{\rm succ})=\tr{\rho_\theta M_{\hat\theta}}$, and note that we have assumed a flat prior probability for $\rho_\theta=U_\theta\rho U_\theta^\dagger$. 
Obviously, $F\le F_{\rm av}$. We just need to show that the opposite inequality also~hold.

It is known that in order to maximize $F_{\rm av}$ one can choose a covariant measurement, so that $M_{\hat\theta}=U_{\hat\theta}\Omega U^\dagger_{\hat\theta}$ for a given seed~$\Omega$. 
Because of covariance, we note that for any phase $\theta'$ we have $\tr{\rho_\theta M_{\hat\theta}}=\tr{\rho_{\theta'} M_{\hat\theta+\Delta\theta}}$, where $\Delta\theta=\theta'-\theta$, thus $p(\hat\theta|\theta,{\rm succ})=p(\hat\theta+\Delta\theta|\theta',{\rm succ})$. Likewise, we have $f(\theta,\hat\theta)=f(\theta',\hat\theta+\Delta\theta)$.
By shifting variables $\hat\theta+\Delta\theta\to \hat\theta$ in Eq.~(\ref{Fav}), the integrant becomes independent of the variable $\theta$, which can be trivially integrated  to give 
\begin{equation}
F_{\rm av}= \int d\hat\theta \,p(\hat\theta|\theta',{\rm succ}) f(\theta',\hat\theta)
\end{equation}
for any $\theta'$. It follows that
\begin{equation}
F_{\rm av}=\inf_{\theta\in(-\pi,\pi]} \int d\hat\theta \,p(\hat\theta|\theta,{\rm succ}) f(\theta,\hat\theta)\le F,
\end{equation}
where the last inequality states that the measurement that maximizes  $F_{\rm av}$ need not maximize the worst-case fidelity $F$. We conclude that $F=F_{\rm av}$.

Proceeding along the same lines, we note that
\begin{equation}
S_{\rm av}=\int {d\theta\over2\pi} \int d\hat\theta \,p(\hat\theta|\theta,{\rm succ}) 
=\int d\hat\theta \,p(\hat\theta|\theta',{\rm succ})
\end{equation}
for any $\theta'$. Hence $S_{\rm av}=S$.

We point out here that most of the cited work dealing with quantum metrology, rather than the minimax approach used here, follows a \emph{pointwise} approach that is aimed at improving the phase sensitivity around a rough estimate of $\theta$. This very powerful and general approach is based on the quantum Cramer-Rao bound~\cite{braunstein_statistical_1994}, and  one can often argue that the so-obtained sensitivity can be attained by a suitable two-step adaptive protocol. However, the working hypotheses and the Cramer-Rao bound entail some  subtleties that  are often ignored, which can lead to erroneous conclusions  \cite{berry_optimal_2012,giovannetti_sub-heisenberg_2012}, wrong bounds, or misleading accounting of resources, even in the asymptotic regime of many such resources (see for instance \cite{hayashi_comparison_2011}). In the particular case of probabilistic metrology the direct application of the pointwise approach can lead to unphysical results, as pointed out in~\cite{chiribella_quantum_2013}. 
Here, we  follow a \emph{global approach} \cite{holevo_probabilistic_1982,aspachs_phase_2009,kolodynski_phase_2010} where no a priori knowledge about the phase is assumed, so instead the phase $\theta$ takes random equidistributed values in the interval $(-\pi,\pi]$.
Within this approach
the allocation of resources is straight-forward and the results are valid both for asymptotically large and finite amount of resources.

\medskip

\noindent{\bf The continuum limit: Particle in a potential box.}
Proceeding as in~\cite{gendra_quantum_2013,gendra_optimal_2013}, one can easily derive from Eqs.~(\ref{eq:fidj}), (\ref{eq:H^j_mm}) and~(\ref{eq:fidjmax}) the following equation:
\begin{eqnarray}
\langle\xi^j|H^j\!|\xi^j\rangle&=&\!\sum_{m=-j}^{j-1} \!\!\left\{\!\left(\xi_{m+1}\!-\!\xi_m\right)^2\!\!+{V^j_m\over j^2}\xi^2_m\!\right\}\nonumber\\
&+& a_{-j}\xi^2_{-j}\!+a_j\xi^2_{j},
\label{eq:discrete}
\end{eqnarray}

where $V^j_m=2j^2(1-a_m)$ and we have dropped the superscript $j$ in $\xi^j_m$ to simplify the expression. In the asymptotic limit, as $j$ becomes very large, $m/j=x$ approaches a continuum variable that takes values in the interval~$[-1,1]$. Accordingly, the values $\{\sqrt j \xi_m\}$ approach a real function that we denote by $\varphi(x)$. With this, the former equation becomes 
\begin{eqnarray}
\label{eq: ?}
 \langle\xi^j|H^j\!|\xi^j\rangle\!&=&{
1\over j^2}%
\int_{-1}^1 \!\!dx \left\{  \left[{d\varphi(x)\over dx}\right]^2\!\!+{V^j(x)}\varphi(x)^{2}\right\},\nonumber\\[.5em]
:\!&=&{
1\over j^2}\bra{\varphi} {\cal H}^j\ket{\varphi},
\end{eqnarray}
where
\begin{equation}
V^j(x)= 2 j^2 (1-a^j_{xj}),\quad {\cal H}^j:=-d^2/dx^2+V^j(x),
\end{equation}
and we have dropped the boundary term $[\varphi^2(-1)+\varphi^2(-1)]/j$ that stems from the second line in~(\ref{eq:discrete}). Minimization of~$\langle\xi^j|H^j\!|\xi^j\rangle$ require the vanishing of this term, and Eq.~(\ref{eq: sigmaj2}) readily follows.
The formula
\begin{equation}
V^j(x)= j{1-r^2\over2r\sqrt{1-(1-r^2)x^2}} 
\end{equation}
[also in Eq.~(\ref{eq:pot})] follows from the asymptotic expression of~$a^j_m$, defined in~(\ref{eq:H^j_mm}). Our starting point is Eq.~(\ref{a^j_m D}) and~(\ref{eq:CG}). The sum over $k$ in the latter can be evaluated using the Euler–Maclaurin formula and the saddle point approximation.

\medskip

\noindent{\bf Symmetric probe is optimal and no benefit in probe-ancilla entanglement.}
We next show that permutation invariance enables us to choose with  no loss of generality the probe state  $|\psi\rangle$ from the symmetric subspace ${\cal H}^{\otimes n}_+$ and the seed~$\Omega$ to be fully symmetric. 

We first write Eqs.~(\ref{F(S) b}) and~(\ref{S b}) in a more compact form. We define $\Delta$ as the non-trivial term of the fidelity through the relation $F(S)=(1+S^{-1}\max_{\psi,\Omega}\Delta)/2$, where the maximization is performed also over the probe states since here we are concerned with the ultimate precision bound. We also introduce a slight modification of ${\cal D}$ that includes the Kronecker delta tensor:  ${\cal D}_l:=\sum_{b,b'} r^{|b+b'|}\delta_{|b'|,|b|+l}|b\rangle\langle b'|$, $l=0,1$. Then,
\begin{equation}
\Delta={\rm tr}\left[(\psi\circ \Omega) {\cal D}_1\right],\quad
S={\rm tr}\left[(\psi\circ \Omega) {\cal D}_0\right],
\label{Delta S}
\end{equation}
where we have used that ${\rm tr}[A(B\circ C)]={\rm tr}[(C\circ A)B]$ if $B=B^t$.
The result we wish to show follows from the invariance of the noise under permutations of the $n$ qubits, namely, from $U_\pi {\cal D}_lU^\dagger_\pi={\cal D}_l$, for any $\pi\in {S}_n$, which implies that the very same value of $\Delta$ and $S$ attained by some given measurement seed $\Omega$ and some  initial state $\psi$, i.e., attained by $\psi\circ\Omega$, can also be attained by $U_\pi( \psi\circ\Omega) U^\dagger_\pi$, and likewise by the average $(n!)^{-1}\sum_{\pi\in {S}_n}U_{\pi}( \psi\circ\Omega) U^\dagger_{\pi}$.

The proof starts with yet a few more definitions: given a fully general probe state  $|\psi\rangle$, we define the $n+1$ normalized states 
$|\phi_\beta\rangle=\sum_{b\in B_\beta}(\psi_b/\psi_\beta)|b\rangle$, $\beta=0,1,\dots, n$ where 
\mbox{$\psi^2_\beta=\sum_{b\in B_\beta}|\psi_b|^2$}, and write
$
|\psi\rangle=\sum_{\beta=0}^n \psi_\beta |\phi_\beta\rangle
$. 
Additionally, we define
\begin{equation}
|\phi\rangle=\sum_{\beta=0}^n
{n\choose\beta}^{1/2}
|\phi_\beta\rangle, \quad \phi=|\phi\rangle\langle\phi|  .
\end{equation}
We obviously have $\phi\circ\Omega\ge0$, as the Hadamard product of two positive operators is also a positive operator, and
\mbox{$
(n!)^{-1}\sum_{\pi\in{S}_n} U_\pi \left(\phi\circ\Omega\right) U^\dagger_\pi\ge 0
$}, as this expression is a convex combination of positive operators.
Similarly, the seed condition~$\openone-\Omega\ge0$ implies $(n!)^{-1}\sum_{\pi\in{S}_n} U_\pi \left[\phi\circ\left(\openone-\Omega\right)\right] U^\dagger_\pi\ge 0$. But
\begin{equation}
{1\over n!}\!\!\sum_{\pi\in{S}_n} \!\!U_\pi\! \left(\phi\circ \openone\right)\! U^\dagger_\pi
\!=\!\sum_{\beta=0}^n {{n\choose\beta}\over n!}\!
\left(
\sum_{\pi\in{S}_n} \!U_\pi\phi_\beta U^\dagger_\pi  \right)   \circ \openone,
\label{<1a}
\end{equation}
since the diagonal entries of $\phi$ and $\phi_\beta$ transform among themselves under permutations. 
The right hand side can be written as
\begin{equation}
\sum_{\beta=0}^n\sum_{b\in B_\beta}\!\!
 {{n \choose \beta}\over n!}\!\!\sum_{\pi\in{S}_n}\!\! \!{|\psi_{\pi^{-1}\!(b)}|^2\kern-.3em\over\psi^2_\beta} |b\rangle\langle b|
\!=\!
\sum_{\beta=0}^n\sum_{b\in B_\beta} \!\!
|b\rangle\langle b|
=\openone,
\label{<1}
\end{equation}

where we have used that,  for any  $b\in B_\beta$, the set $\{\pi^{-1}(b)\}_{\pi\in{S}_n}$ contains exactly~$\beta!(n-\beta)!$ times each one of the elements of $B_\beta$. 
It follows from Eqs.~(\ref{<1a})  and~(\ref{<1}) that $ \Omega^{\rm sym}:=(n!)^{-1}\sum_{\pi\in{S}_n} U_\pi \left(\phi\circ\Omega\right) U^\dagger_\pi$ satisfies
$0\le \Omega^{\rm sym}\le \openone$ and is invariant under permutations of the $n$ qubits. It is, therefore, a legitimate fully symmetric measurement seed. Moreover,
\begin{equation}
{1\over n!}\!\!\sum_{\pi\in{S}_n}\!\!U_\pi\!\left(\psi\!\circ\!\Omega\right)\!U^\dagger_\pi
\!=\!
\sum_{\beta,\beta'} {\psi_\beta\psi_{\beta'}\over  {n \choose \beta}^{1/2} {n \choose \beta'}^{1/2}}
\openone_\beta
\Omega^{\rm sym}
 \openone_{\beta'},
 \label{symm str}
 \end{equation}
where $\openone_\beta$ is the projector into the subspace with $|b|=\beta$, namely $\openone_\beta:=\sum_{b\in B_\beta}|b\rangle\langle b|$. Thus, recalling the definition of $|\beta\rangle$ in Eq.~(\ref{basis H_+}), the righthand side of~(\ref{symm str}) can be readily written as
\begin{equation}
\Bigg(\sum_{\beta,\beta'} \psi_\beta\psi_{\beta'}|\beta\rangle\langle\beta'|\Bigg)\circ \Omega^{\rm sym}
=\psi^{\rm sym}\circ \Omega^{\rm sym} ,
\end{equation}
where~\mbox{$|\psi^{\rm sym}\rangle:=\sum_{\beta=0}^n\psi_\beta |\beta\rangle\in{\cal H}^{\otimes n}_+$}. It follows from these results and Eq.~(\ref{Delta S}) that the very same fidelity and success probability attained by any pair $(|\psi\rangle,\Omega)$ of probe state and measurement seed is also attained by the state $|\psi^{\rm sym}\rangle\in{\cal H}^{\otimes n}_+$ and the fully symmetric seed $\Omega^{\rm sym}$. This completes the proof.

Now that we have learned that no boost in performance can be achieved by considering probe states more general than those in the symmetric subspace~${\cal H}^{\otimes n}_+$ (in the subspace of maximum  spin $j=J$),
we may wonder if entangling the probe with some ancillary system could enhance the precision. 
Here we show that this possibility can be immediately ruled out, thus extending the generality of our result. For this purpose we take the general probe-ancilla state $\ket{\Psi_{\rm PA}}=\sum_{b} \psi_{b} \ket{b}\ket{\chi_{b}}$,
where $\ket{\chi_{b}}$ are normalized states (not necessarily orthogonal) of the ancillary system.
The action of the phase evolution and noise on the probe leads to a state of the form \mbox{$\rho_{\rm PA}(\theta)\;=\;\sum_{b,b'} r^{|b+b'|} \;\expo{i \theta(|b|-|b'|)} \;\psi_{b}\psi_{b'} \ketbra{b}{b'}\otimes\ketbra{\chi_{b}}{\chi_{b'}}
$}.
This state could as well be prepared without the need of an ancillary system by taking instead an initial  probe state $\ket{\psi}=\sum_{b}\psi_{b}\ket{b}$ and performing the trace-preserving completely positive map defined by~$\ket{b}\to\ket{b}\ket{\chi_{b}}$ before implementing the measurement. This map can, of course, be interpreted as part of the measurement. It would correspond to a particular Neumark dilation of some measurement performed on the probe system alone, and hence it is included in our analysis.

\medskip

\noindent{\bf Scavenging at the ultimate precision limit.}
The gentle measurement lemma \cite{winter_coding_1999} states that if measurement outcome occurs with a very high probability, e.g., an unfavorable event in some probabilistic protocol, which happens with probability~$\bar{S}=\mathrm{tr}[\mathcal{\bar {\mathscr F}}(\rho_{\theta})]=1-\epsilon$, then, in that event, the measurement causes very  little disturbance to the state namely $\parallel\rho_{\theta}-\bar\rho_{\theta}\parallel_{1}\leq \sqrt{2}\epsilon$, where $\bar{\rho}={\bar{\mathscr F}}(\rho_{\theta})/{\bar S}$.

Indeed, from  \eqref{F_max}, we find that  the fidelity of the scavenged events, $\bar F(S)$, rapidly approaches that of the optimal deterministic machine $F(S=0)$:
\begin{eqnarray}
&&F(0)\!-\!\bar F(S)\!=\!\!\!\max_{0\leq \Omega\leq \id}\!\! \tr{W\!\rho_{\theta}}-\!\!\!\max_{0\leq \bar\Omega\leq \id}\!\! \tr{\bar W\!\bar\rho_{\theta}}\nonumber\\
&&\leq\min_{0\leq \Omega\leq \id} \mathrm{tr}[W(\rho_{\theta}\!-\!\bar\rho_{\theta})]\leq \parallel\!\rho_{\theta}\!-\!\bar\rho_{\theta}\!\parallel_{1}
\leq\! \sqrt{2}{S},
\label{eq:gentle}
\end{eqnarray}
where $W$ (likewise $\bar W$) is shorthand for the matrix with entries $W_{b,b'}=\Omega_{b,b'}\delta_{|b|,|b'|+1}$. 

We recall that, as $F(S)$ approaches the ultimate bound $F_{\rm ult}=1-(1-r^2)/(4nr)$, the success probability $S$ decreases exponentially. Eq.~(\ref{eq:gentle}) thus shows that such likely failure is not ruinous since 
in that event one can still recover the deterministic bound, i.e.,  one has \mbox{$\bar{F}=F(0)=1-(1-r^2)/(4nr^2)$}.

\section{Acknowledgements}

We acknowledge financial support from ERDF: European Regional Development Fund. This research was supported by  the Spanish MICINN, through contract FIS2008-01236 and the Generalitat de
Catalunya CIRIT, contract  2009SGR-0985.


\begin{thebibliography}{10}
\expandafter\ifx\csname url\endcsname\relax
  \def\url#1{\texttt{#1}}\fi
\expandafter\ifx\csname urlprefix\endcsname\relax\def\urlprefix{URL }\fi
\providecommand{\bibinfo}[2]{#2}
\providecommand{\eprint}[2][]{\url{#2}}

\bibitem{dowling_quantum_2003}
\bibinfo{author}{Dowling, J.~P.} \& \bibinfo{author}{Milburn, G.~J.}
\newblock \bibinfo{title}{Quantum technology: the second quantum revolution}.
\newblock \emph{\bibinfo{journal}{Philos. Trans. R. Soc. London, Ser. A}}
  \textbf{\bibinfo{volume}{361}}, \bibinfo{pages}{1655--1674}
  (\bibinfo{year}{2003}).

\bibitem{slavik_all-optical_2010}
\bibinfo{author}{Slav{\'\i}k, R.} \emph{et~al.}
\newblock \bibinfo{title}{All-optical phase and amplitude regenerator for
  next-generation telecommunications systems}.
\newblock \emph{\bibinfo{journal}{Nat. Photon.}} \textbf{\bibinfo{volume}{4}},
  \bibinfo{pages}{690--695} (\bibinfo{year}{2010}).

\bibitem{chen_optical_2012}
\bibinfo{author}{Chen, J.}, \bibinfo{author}{Habif, J.~L.},
  \bibinfo{author}{Dutton, Z.}, \bibinfo{author}{Lazarus, R.} \&
  \bibinfo{author}{Guha, S.}
\newblock \bibinfo{title}{Optical codeword demodulation with error rates below
  the standard quantum limit using a conditional nulling receiver}.
\newblock \emph{\bibinfo{journal}{Nat. Photon.}} \textbf{\bibinfo{volume}{6}},
  \bibinfo{pages}{374--379} (\bibinfo{year}{2012}).

\bibitem{inoue_differential_2002}
\bibinfo{author}{Inoue, K.}, \bibinfo{author}{Waks, E.} \&
  \bibinfo{author}{Yamamoto, Y.}
\newblock \bibinfo{title}{Differential phase shift quantum key distribution}.
\newblock \emph{\bibinfo{journal}{Phys. Rev. Lett.}}
  \textbf{\bibinfo{volume}{89}}, \bibinfo{pages}{037902}
  (\bibinfo{year}{2002}).

\bibitem{budker_optical_2007}
\bibinfo{author}{Budker, D.}
\newblock \bibinfo{title}{Optical magnetometry}.
\newblock \emph{\bibinfo{journal}{Nat. Phys.}} \textbf{\bibinfo{volume}{3}},
  \bibinfo{pages}{227--234} (\bibinfo{year}{2007}).

\bibitem{napolitano_interaction-based_2011}
\bibinfo{author}{Napolitano, M.} \emph{et~al.}
\newblock \bibinfo{title}{Interaction-based quantum metrology showing scaling
  beyond the heisenberg limit}.
\newblock \emph{\bibinfo{journal}{Nature}} \textbf{\bibinfo{volume}{471}},
  \bibinfo{pages}{486--489} (\bibinfo{year}{2011}).

\bibitem{taylor_biological_2013}
\bibinfo{author}{Taylor, M.~A.} \emph{et~al.}
\newblock \bibinfo{title}{Biological measurement beyond the quantum limit}.
\newblock \emph{\bibinfo{journal}{Nat. Photon.}} \textbf{\bibinfo{volume}{7}},
  \bibinfo{pages}{229--233} (\bibinfo{year}{2013}).

\bibitem{crespi_measuring_2012}
\bibinfo{author}{Crespi, A.} \emph{et~al.}
\newblock \bibinfo{title}{Measuring protein concentration with entangled
  photons}.
\newblock \emph{\bibinfo{journal}{Appl. Phys. Lett.}}
  \textbf{\bibinfo{volume}{100}}, \bibinfo{pages}{233704}
  (\bibinfo{year}{2012}).

\bibitem{caves_quantum-mechanical_1981}
\bibinfo{author}{Caves, C.~M.}
\newblock \bibinfo{title}{Quantum-mechanical noise in an interferometer}.
\newblock \emph{\bibinfo{journal}{Phys. Rev. D}} \textbf{\bibinfo{volume}{23}},
  \bibinfo{pages}{1693--1708} (\bibinfo{year}{1981}).

\bibitem{collaboration_gravitational_2011}
\bibinfo{author}{Collaboration, T. L.~S.}
\newblock \bibinfo{title}{A gravitational wave observatory operating beyond the
  quantum shot-noise limit}.
\newblock \emph{\bibinfo{journal}{Nat. Phys.}} \textbf{\bibinfo{volume}{7}},
  \bibinfo{pages}{962--965} (\bibinfo{year}{2011}).

\bibitem{bollinger_optimal_1996}
\bibinfo{author}{Bollinger, J.~J.}, \bibinfo{author}{Itano, W.~M.},
  \bibinfo{author}{Wineland, D.~J.} \& \bibinfo{author}{Heinzen, D.~J.}
\newblock \bibinfo{title}{Optimal frequency measurements with maximally
  correlated states}.
\newblock \emph{\bibinfo{journal}{Phys. Rev. A}} \textbf{\bibinfo{volume}{54}},
  \bibinfo{pages}{R4649--R4652} (\bibinfo{year}{1996}).

\bibitem{huelga_improvement_1997}
\bibinfo{author}{Huelga, S.~F.} \emph{et~al.}
\newblock \bibinfo{title}{Improvement of frequency standards with quantum
  entanglement}.
\newblock \emph{\bibinfo{journal}{Phys. Rev. Lett.}}
  \textbf{\bibinfo{volume}{79}}, \bibinfo{pages}{3865--3868}
  (\bibinfo{year}{1997}).

\bibitem{wineland_spin_1992}
\bibinfo{author}{Wineland, D.~J.}, \bibinfo{author}{Bollinger, J.~J.},
  \bibinfo{author}{Itano, W.~M.}, \bibinfo{author}{Moore, F.~L.} \&
  \bibinfo{author}{Heinzen, D.~J.}
\newblock \bibinfo{title}{Spin squeezing and reduced quantum noise in
  spectroscopy}.
\newblock \emph{\bibinfo{journal}{Phys. Rev. A}} \textbf{\bibinfo{volume}{46}},
  \bibinfo{pages}{R6797--R6800} (\bibinfo{year}{1992}).

\bibitem{komar_quantum_2014}
\bibinfo{author}{K{\'o}m{\'a}r, P.} \emph{et~al.}
\newblock \bibinfo{title}{A quantum network of clocks}.
\newblock \emph{\bibinfo{journal}{Nat. Phys.}}  (\bibinfo{year}{2014}).

\bibitem{giovannetti_quantum_2006}
\bibinfo{author}{Giovannetti, V.}, \bibinfo{author}{Lloyd, S.} \&
  \bibinfo{author}{Maccone, L.}
\newblock \bibinfo{title}{Quantum metrology}.
\newblock \emph{\bibinfo{journal}{Phys. Rev. Lett.}}
  \textbf{\bibinfo{volume}{96}}, \bibinfo{pages}{010401}
  (\bibinfo{year}{2006}).

\bibitem{giovannetti_advances_2011}
\bibinfo{author}{Giovannetti, V.}, \bibinfo{author}{Lloyd, S.} \&
  \bibinfo{author}{Maccone, L.}
\newblock \bibinfo{title}{Advances in quantum metrology}.
\newblock \emph{\bibinfo{journal}{Nat. Photon.}} \textbf{\bibinfo{volume}{5}},
  \bibinfo{pages}{222--229} (\bibinfo{year}{2011}).

\bibitem{demkowicz-dobrzanski_elusive_2012}
\bibinfo{author}{Demkowicz-Dobrza{\'n}ski, R.},
  \bibinfo{author}{Ko{\l}ody{\'n}ski, J.} \& \bibinfo{author}{Gu{\c t}{\u a},
  M.}
\newblock \bibinfo{title}{The elusive heisenberg limit in quantum-enhanced
  metrology}.
\newblock \emph{\bibinfo{journal}{Nat. Commun.}} \textbf{\bibinfo{volume}{3}},
  \bibinfo{pages}{1063} (\bibinfo{year}{2012}).

\bibitem{escher_general_2011}
\bibinfo{author}{Escher, B.~M.}, \bibinfo{author}{de~Matos~Filho, R.~L.} \&
  \bibinfo{author}{Davidovich, L.}
\newblock \bibinfo{title}{General framework for estimating the ultimate
  precision limit in noisy quantum-enhanced metrology}.
\newblock \emph{\bibinfo{journal}{Nat. Phys.}} \textbf{\bibinfo{volume}{7}},
  \bibinfo{pages}{406--411} (\bibinfo{year}{2011}).

\bibitem{knysh_scaling_2011}
\bibinfo{author}{Knysh, S.}, \bibinfo{author}{Smelyanskiy, V.~N.} \&
  \bibinfo{author}{Durkin, G.~A.}
\newblock \bibinfo{title}{Scaling laws for precision in quantum interferometry
  and the bifurcation landscape of the optimal state}.
\newblock \emph{\bibinfo{journal}{Phys. Rev. A}} \textbf{\bibinfo{volume}{83}},
  \bibinfo{pages}{021804} (\bibinfo{year}{2011}).

\bibitem{aspachs_phase_2009}
\bibinfo{author}{Aspachs, M.}, \bibinfo{author}{Calsamiglia, J.},
  \bibinfo{author}{Mu{\~n}oz-Tapia, R.} \& \bibinfo{author}{Bagan, E.}
\newblock \bibinfo{title}{Phase estimation for thermal gaussian states}.
\newblock \emph{\bibinfo{journal}{Phys. Rev. A}} \textbf{\bibinfo{volume}{79}},
  \bibinfo{pages}{033834} (\bibinfo{year}{2009}).

\bibitem{chaves_noisy_2013}
\bibinfo{author}{Chaves, R.}, \bibinfo{author}{Brask, J.~B.},
  \bibinfo{author}{Markiewicz, M.}, \bibinfo{author}{Ko{\l}ody{\'n}ski, J.} \&
  \bibinfo{author}{Ac{\'\i}n, A.}
\newblock \bibinfo{title}{Noisy metrology beyond the standard quantum limit}.
\newblock \emph{\bibinfo{journal}{Phys. Rev. Lett.}}
  \textbf{\bibinfo{volume}{111}}, \bibinfo{pages}{120401}
  (\bibinfo{year}{2013}).

\bibitem{chin_quantum_2012}
\bibinfo{author}{Chin, A.~W.}, \bibinfo{author}{Huelga, S.~F.} \&
  \bibinfo{author}{Plenio, M.~B.}
\newblock \bibinfo{title}{Quantum metrology in non-markovian environments}.
\newblock \emph{\bibinfo{journal}{Phys. Rev. Lett.}}
  \textbf{\bibinfo{volume}{109}}, \bibinfo{pages}{233601}
  (\bibinfo{year}{2012}).

\bibitem{jeske_quantum_2013}
\bibinfo{author}{Jeske, J.}, \bibinfo{author}{Cole, J.~H.} \&
  \bibinfo{author}{Huelga, S.}
\newblock \bibinfo{title}{Quantum metrology in the presence of spatially
  correlated noise: Restoring heisenberg scaling}.
\newblock \emph{\bibinfo{journal}{arXiv:1307.6301}}  (\bibinfo{year}{2013}).

\bibitem{ostermann_protected_2013}
\bibinfo{author}{Ostermann, L.}, \bibinfo{author}{Ritsch, H.} \&
  \bibinfo{author}{Genes, C.}
\newblock \bibinfo{title}{Protected state enhanced quantum metrology with
  interacting two-level ensembles}.
\newblock \emph{\bibinfo{journal}{Phys. Rev. Lett.}}
  \textbf{\bibinfo{volume}{111}}, \bibinfo{pages}{123601}
  (\bibinfo{year}{2013}).

\bibitem{szankowski_parameter_2012}
\bibinfo{author}{Szankowski, P.}, \bibinfo{author}{Trippenbach, M.} \&
  \bibinfo{author}{Chwedenczuk, J.}
\newblock \bibinfo{title}{Parameter estimation in a memory-assisted noisy
  quantum interferometry}.
\newblock \emph{\bibinfo{journal}{arXiv:1212.2528}}  (\bibinfo{year}{2012}).

\bibitem{boixo_generalized_2007}
\bibinfo{author}{Boixo, S.}, \bibinfo{author}{Flammia, S.~T.},
  \bibinfo{author}{Caves, C.~M.} \& \bibinfo{author}{Geremia, J.}
\newblock \bibinfo{title}{Generalized limits for single-parameter quantum
  estimation}.
\newblock \emph{\bibinfo{journal}{Phys. Rev. Lett.}}
  \textbf{\bibinfo{volume}{98}}, \bibinfo{pages}{090401}
  (\bibinfo{year}{2007}).

\bibitem{dur_improved_2014}
\bibinfo{author}{D{\"u}r, W.}, \bibinfo{author}{Skotiniotis, M.},
  \bibinfo{author}{Fr{\"o}wis, F.} \& \bibinfo{author}{Kraus, B.}
\newblock \bibinfo{title}{Improved quantum metrology using quantum error
  correction}.
\newblock \emph{\bibinfo{journal}{Phys. Rev. Lett.}}
  \textbf{\bibinfo{volume}{112}}, \bibinfo{pages}{080801}
  (\bibinfo{year}{2014}).

\bibitem{kessler_quantum_2014}
\bibinfo{author}{Kessler, E.}, \bibinfo{author}{Lovchinsky, I.},
  \bibinfo{author}{Sushkov, A.} \& \bibinfo{author}{Lukin, M.}
\newblock \bibinfo{title}{Quantum error correction for metrology}.
\newblock \emph{\bibinfo{journal}{Phys. Rev. Lett.}}
  \textbf{\bibinfo{volume}{112}}, \bibinfo{pages}{150802}
  (\bibinfo{year}{2014}).

\bibitem{arrad_increasing_2014}
\bibinfo{author}{Arrad, G.}, \bibinfo{author}{Vinkler, Y.},
  \bibinfo{author}{Aharonov, D.} \& \bibinfo{author}{Retzker, A.}
\newblock \bibinfo{title}{Increasing sensing resolution with error correction}.
\newblock \emph{\bibinfo{journal}{Phys. Rev. Lett.}}
  \textbf{\bibinfo{volume}{112}}, \bibinfo{pages}{150801}
  (\bibinfo{year}{2014}).

\bibitem{fiurasek_optimal_2006}
\bibinfo{author}{Fiur{\'a}{\v s}ek, J.}
\newblock \bibinfo{title}{Optimal probabilistic estimation of quantum states}.
\newblock \emph{\bibinfo{journal}{New J. Phys.}} \textbf{\bibinfo{volume}{8}},
  \bibinfo{pages}{192} (\bibinfo{year}{2006}).

\bibitem{gendra_quantum_2013}
\bibinfo{author}{Gendra, B.}, \bibinfo{author}{Ronco-Bonvehi, E.},
  \bibinfo{author}{Calsamiglia, J.}, \bibinfo{author}{Mu{\~n}oz-Tapia, R.} \&
  \bibinfo{author}{Bagan, E.}
\newblock \bibinfo{title}{Quantum metrology assisted by abstention}.
\newblock \emph{\bibinfo{journal}{Phys. Rev. Lett.}}
  \textbf{\bibinfo{volume}{110}}, \bibinfo{pages}{100501}
  (\bibinfo{year}{2013}).

\bibitem{gendra_optimal_2013}
\bibinfo{author}{Gendra, B.}, \bibinfo{author}{Ronco-Bonvehi, E.},
  \bibinfo{author}{Calsamiglia, J.}, \bibinfo{author}{Mu{\~n}oz-Tapia, R.} \&
  \bibinfo{author}{Bagan, E.}
\newblock \bibinfo{title}{Optimal parameter estimation with a fixed rate of
  abstention}.
\newblock \emph{\bibinfo{journal}{Phys. Rev. A}} \textbf{\bibinfo{volume}{88}},
  \bibinfo{pages}{012128} (\bibinfo{year}{2013}).

\bibitem{marek_optimal_2013}
\bibinfo{author}{Marek, P.}
\newblock \bibinfo{title}{Optimal probabilistic measurement of phase}.
\newblock \emph{\bibinfo{journal}{Phys. Rev. A}} \textbf{\bibinfo{volume}{88}},
  \bibinfo{pages}{045802} (\bibinfo{year}{2013}).

\bibitem{Note1}
\bibinfo{note}{The analogy with the quantum mechanical problem of the
  groundstate of a particle in a potential was first presented by J.
  Calsamiglia, Abstention-enhanced metrology, Noise Information \& Complexity @
  Quantum Scale, Ettore Majorana Centre, Erice, Italy (2013). A similar
  equivalence has been drawn recently in \cite {knysh_true_2014} in a
  point-wise approach (see Methods)}.

\bibitem{combes_quantum_2014}
\bibinfo{author}{Combes, J.}, \bibinfo{author}{Ferrie, C.},
  \bibinfo{author}{Jiang, Z.} \& \bibinfo{author}{Caves, C.~M.}
\newblock \bibinfo{title}{Quantum limits on postselected, probabilistic quantum
  metrology}.
\newblock \emph{\bibinfo{journal}{Phys. Rev. A}} \textbf{\bibinfo{volume}{89}},
  \bibinfo{pages}{052117} (\bibinfo{year}{2014}).

\bibitem{Note2}
\bibinfo{note}{In our approach we take the total number of subsystems as a
  resource. This is in contrast to other (usually point-wise) approaches where
  the number of subsystems in a single run is accounted, but an unlimited
  number of repetitions of each run is allowed. See Methods.}

\bibitem{bagan_optimal_2012}
\bibinfo{author}{Bagan, E.}, \bibinfo{author}{Mu{\~n}oz-Tapia, R.},
  \bibinfo{author}{Olivares-Renter{\'\i}a, G.~A.} \& \bibinfo{author}{Bergou,
  J.~A.}
\newblock \bibinfo{title}{Optimal discrimination of quantum states with a fixed
  rate of inconclusive outcomes}.
\newblock \emph{\bibinfo{journal}{Phys. Rev. A}} \textbf{\bibinfo{volume}{86}},
  \bibinfo{pages}{040303} (\bibinfo{year}{2012}).

\bibitem{fiurasek_optimal_2003}
\bibinfo{author}{Fiur{\'a}{\v s}ek, J.} \& \bibinfo{author}{Je{\v z}ek, M.}
\newblock \bibinfo{title}{Optimal discrimination of mixed quantum states
  involving inconclusive results}.
\newblock \emph{\bibinfo{journal}{Phys. Rev. A}} \textbf{\bibinfo{volume}{67}},
  \bibinfo{pages}{012321} (\bibinfo{year}{2003}).

\bibitem{lutkenhaus_bell_1999}
\bibinfo{author}{L{\"u}tkenhaus, N.}, \bibinfo{author}{Calsamiglia, J.} \&
  \bibinfo{author}{Suominen, K.-A.}
\newblock \bibinfo{title}{Bell measurements for teleportation}.
\newblock \emph{\bibinfo{journal}{Phys. Rev. A}} \textbf{\bibinfo{volume}{59}},
  \bibinfo{pages}{3295--3300} (\bibinfo{year}{1999}).

\bibitem{knill_scheme_2001}
\bibinfo{author}{Knill, E.}, \bibinfo{author}{Laflamme, R.} \&
  \bibinfo{author}{Milburn, G.~J.}
\newblock \bibinfo{title}{A scheme for efficient quantum computation with
  linear optics}.
\newblock \emph{\bibinfo{journal}{Nature}} \textbf{\bibinfo{volume}{409}},
  \bibinfo{pages}{46--52} (\bibinfo{year}{2001}).

\bibitem{ambainis_exact_2014}
\bibinfo{author}{Ambainis, A.}, \bibinfo{author}{Gruska, J.} \&
  \bibinfo{author}{Zheng, S.}
\newblock \bibinfo{title}{Exact query complexity of some special classes of
  boolean functions}.
\newblock \emph{\bibinfo{journal}{arXiv:1404.1684}}  (\bibinfo{year}{2014}).

\bibitem{montanaro_exact_2013}
\bibinfo{author}{Montanaro, A.}, \bibinfo{author}{Jozsa, R.} \&
  \bibinfo{author}{Mitchison, G.}
\newblock \bibinfo{title}{On exact quantum query complexity}.
\newblock \emph{\bibinfo{journal}{Algorithmica}} \bibinfo{pages}{1--22}
  (\bibinfo{year}{2013}).

\bibitem{ferreyrol_implementation_2010}
\bibinfo{author}{Ferreyrol, F.} \emph{et~al.}
\newblock \bibinfo{title}{Implementation of a nondeterministic optical
  noiseless amplifier}.
\newblock \emph{\bibinfo{journal}{Phys. Rev. Lett.}}
  \textbf{\bibinfo{volume}{104}}, \bibinfo{pages}{123603}
  (\bibinfo{year}{2010}).

\bibitem{xiang_entanglement-enhanced_2011}
\bibinfo{author}{Xiang, G.~Y.}, \bibinfo{author}{Higgins, B.~L.},
  \bibinfo{author}{Berry, D.~W.}, \bibinfo{author}{Wiseman, H.~M.} \&
  \bibinfo{author}{Pryde, G.~J.}
\newblock \bibinfo{title}{Entanglement-enhanced measurement of a completely
  unknown optical phase}.
\newblock \emph{\bibinfo{journal}{Nat. Photon.}} \textbf{\bibinfo{volume}{5}},
  \bibinfo{pages}{43--47} (\bibinfo{year}{2011}).

\bibitem{zavattaa._high-fidelity_2011}
\bibinfo{author}{{Zavatta, A.}}, \bibinfo{author}{{Fiurasek, J.}} \&
  \bibinfo{author}{{Bellini, M.}}
\newblock \bibinfo{title}{A high-fidelity noiseless amplifier for quantum light
  states}.
\newblock \emph{\bibinfo{journal}{Nat. Photon.}} \textbf{\bibinfo{volume}{5}},
  \bibinfo{pages}{52--60} (\bibinfo{year}{2011}).

\bibitem{kocsis_heralded_2013}
\bibinfo{author}{Kocsis, S.}, \bibinfo{author}{Xiang, G.~Y.},
  \bibinfo{author}{Ralph, T.~C.} \& \bibinfo{author}{Pryde, G.~J.}
\newblock \bibinfo{title}{Heralded noiseless amplification of a photon
  polarization qubit}.
\newblock \emph{\bibinfo{journal}{Nat. Phys.}} \textbf{\bibinfo{volume}{9}},
  \bibinfo{pages}{23--28} (\bibinfo{year}{2013}).

\bibitem{chiribella_optimal_2013}
\bibinfo{author}{Chiribella, G.} \& \bibinfo{author}{Xie, J.}
\newblock \bibinfo{title}{Optimal design and quantum benchmarks for coherent
  state amplifiers}.
\newblock \emph{\bibinfo{journal}{Phys. Rev. Lett.}}
  \textbf{\bibinfo{volume}{110}}, \bibinfo{pages}{213602}
  (\bibinfo{year}{2013}).

\bibitem{dressel_colloquium:_2014}
\bibinfo{author}{Dressel, J.}, \bibinfo{author}{Malik, M.},
  \bibinfo{author}{Miatto, F.~M.}, \bibinfo{author}{Jordan, A.~N.} \&
  \bibinfo{author}{Boyd, R.~W.}
\newblock \bibinfo{title}{Colloquium: Understanding quantum weak values: Basics
  and applications}.
\newblock \emph{\bibinfo{journal}{Rev. Mod. Phys.}}
  \textbf{\bibinfo{volume}{86}}, \bibinfo{pages}{307--316}
  (\bibinfo{year}{2014}).

\bibitem{hosten_observation_2008}
\bibinfo{author}{Hosten, O.} \& \bibinfo{author}{Kwiat, P.}
\newblock \bibinfo{title}{Observation of the spin hall effect of light via weak
  measurements}.
\newblock \emph{\bibinfo{journal}{Science}} \textbf{\bibinfo{volume}{319}},
  \bibinfo{pages}{787--790} (\bibinfo{year}{2008}).

\bibitem{brunner_measuring_2010}
\bibinfo{author}{Brunner, N.} \& \bibinfo{author}{Simon, C.}
\newblock \bibinfo{title}{Measuring small longitudinal phase shifts: Weak
  measurements or standard interferometry?}
\newblock \emph{\bibinfo{journal}{Phys. Rev. Lett.}}
  \textbf{\bibinfo{volume}{105}}, \bibinfo{pages}{010405}
  (\bibinfo{year}{2010}).

\bibitem{zilberberg_charge_2011}
\bibinfo{author}{Zilberberg, O.}, \bibinfo{author}{Romito, A.} \&
  \bibinfo{author}{Gefen, Y.}
\newblock \bibinfo{title}{Charge sensing amplification via weak values
  measurement}.
\newblock \emph{\bibinfo{journal}{Phys. Rev. Lett.}}
  \textbf{\bibinfo{volume}{106}}, \bibinfo{pages}{080405}
  (\bibinfo{year}{2011}).

\bibitem{jordan_technical_2014}
\bibinfo{author}{Jordan, A.~N.}, \bibinfo{author}{Mart{\'\i}nez-Rinc{\'o}n, J.}
  \& \bibinfo{author}{Howell, J.~C.}
\newblock \bibinfo{title}{Technical advantages for weak-value amplification:
  When less is more}.
\newblock \emph{\bibinfo{journal}{Physical Review X}}
  \textbf{\bibinfo{volume}{4}}, \bibinfo{pages}{011031} (\bibinfo{year}{2014}).

\bibitem{calsamiglia_quantum_2014}
\bibinfo{author}{Calsamiglia, J.}
\newblock \bibinfo{title}{Quantum information: The occasional super
  clock-cloner}.
\newblock \emph{\bibinfo{journal}{Nat. Phys.}} \textbf{\bibinfo{volume}{10}},
  \bibinfo{pages}{91--92} (\bibinfo{year}{2014}).

\bibitem{aaronson_quantum_2005}
\bibinfo{author}{Aaronson, S.}
\newblock \bibinfo{title}{Quantum computing, postselection, and probabilistic
  polynomial-time}.
\newblock \emph{\bibinfo{journal}{Proc. R. Soc. A}}
  \textbf{\bibinfo{volume}{461}}, \bibinfo{pages}{3473--3482}
  (\bibinfo{year}{2005}).

\bibitem{berry_optimal_2012}
\bibinfo{author}{Berry, D.~W.}, \bibinfo{author}{Hall, M. J.~W.},
  \bibinfo{author}{Zwierz, M.} \& \bibinfo{author}{Wiseman, H.~M.}
\newblock \bibinfo{title}{Optimal heisenberg-style bounds for the average
  performance of arbitrary phase estimates}.
\newblock \emph{\bibinfo{journal}{Phys. Rev. A}} \textbf{\bibinfo{volume}{86}},
  \bibinfo{pages}{053813} (\bibinfo{year}{2012}).

\bibitem{cirac_optimal_1999}
\bibinfo{author}{Cirac, J.~I.}, \bibinfo{author}{Ekert, A.~K.} \&
  \bibinfo{author}{Macchiavello, C.}
\newblock \bibinfo{title}{Optimal purification of single qubits}.
\newblock \emph{\bibinfo{journal}{Phys. Rev. Lett.}}
  \textbf{\bibinfo{volume}{82}}, \bibinfo{pages}{4344} (\bibinfo{year}{1999}).

\bibitem{bagan_optimal_2006}
\bibinfo{author}{Bagan, E.}, \bibinfo{author}{Ballester, M.~A.},
  \bibinfo{author}{Gill, R.~D.}, \bibinfo{author}{Monras, A.} \&
  \bibinfo{author}{Munoz-Tapia, R.}
\newblock \bibinfo{title}{Optimal full estimation of qubit mixed states}.
\newblock \emph{\bibinfo{journal}{Phys. Rev. A}} \textbf{\bibinfo{volume}{73}},
  \bibinfo{pages}{032301} (\bibinfo{year}{2006}).

\bibitem{summy_phase_1990}
\bibinfo{author}{Summy, G.~S.} \& \bibinfo{author}{Pegg, D.~T.}
\newblock \bibinfo{title}{Phase optimized quantum states of light}.
\newblock \emph{\bibinfo{journal}{Opt. Commun.}} \textbf{\bibinfo{volume}{77}},
  \bibinfo{pages}{75--79} (\bibinfo{year}{1990}).

\bibitem{knysh_true_2014}
\bibinfo{author}{Knysh, S.~I.}, \bibinfo{author}{Chen, E.~H.} \&
  \bibinfo{author}{Durkin, G.~A.}
\newblock \bibinfo{title}{True limits to precision via unique quantum probe}.
\newblock \emph{\bibinfo{journal}{arXiv:1402.0495}}  (\bibinfo{year}{2014}).

\bibitem{boyd_convex_2004}
\bibinfo{author}{Boyd, S.} \& \bibinfo{author}{Vandenberghe, L.}
\newblock \emph{\bibinfo{title}{Convex Optimization}}
  (\bibinfo{publisher}{Cambridge University Press}, \bibinfo{address}{New York,
  {NY}, {USA}}, \bibinfo{year}{2004}).

\bibitem{winter_coding_1999}
\bibinfo{author}{Winter, A.}
\newblock \bibinfo{title}{Coding theorem and strong converse for quantum
  channels}.
\newblock \emph{\bibinfo{journal}{{IEEE} Trans. Inf. Theor.}}
  \textbf{\bibinfo{volume}{45}}, \bibinfo{pages}{2481--2485}
  (\bibinfo{year}{1999}).

\bibitem{ferrie_weak_2014}
\bibinfo{author}{Ferrie, C.} \& \bibinfo{author}{Combes, J.}
\newblock \bibinfo{title}{Weak value amplification is suboptimal for estimation
  and detection}.
\newblock \emph{\bibinfo{journal}{Phys. Rev. Lett.}}
  \textbf{\bibinfo{volume}{112}}, \bibinfo{pages}{040406}
  (\bibinfo{year}{2014}).

\bibitem{berry_optimal_2000}
\bibinfo{author}{Berry, D.~W.} \& \bibinfo{author}{Wiseman, H.~M.}
\newblock \bibinfo{title}{Optimal states and almost optimal adaptive
  measurements for quantum interferometry}.
\newblock \emph{\bibinfo{journal}{Phys. Rev. Lett.}}
  \textbf{\bibinfo{volume}{85}}, \bibinfo{pages}{5098--5101}
  (\bibinfo{year}{2000}).

\bibitem{bacon_efficient_2006}
\bibinfo{author}{Bacon, D.}, \bibinfo{author}{Chuang, I.~L.} \&
  \bibinfo{author}{Harrow, A.~W.}
\newblock \bibinfo{title}{Efficient quantum circuits for schur and
  clebsch-gordan transforms}.
\newblock \emph{\bibinfo{journal}{Phys. Rev. Lett.}}
  \textbf{\bibinfo{volume}{97}}, \bibinfo{pages}{170502}
  (\bibinfo{year}{2006}).

\bibitem{calsamiglia_local_2010}
\bibinfo{author}{Calsamiglia, J.}, \bibinfo{author}{de~Vicente, J.~I.},
  \bibinfo{author}{Mu{\~n}oz-Tapia, R.} \& \bibinfo{author}{Bagan, E.}
\newblock \bibinfo{title}{Local discrimination of mixed states}.
\newblock \emph{\bibinfo{journal}{Phys. Rev. Lett.}}
  \textbf{\bibinfo{volume}{105}}, \bibinfo{pages}{080504}
  (\bibinfo{year}{2010}).

\bibitem{braunstein_statistical_1994}
\bibinfo{author}{Braunstein, S.~L.} \& \bibinfo{author}{Caves, C.~M.}
\newblock \bibinfo{title}{Statistical distance and the geometry of quantum
  states}.
\newblock \emph{\bibinfo{journal}{Phys. Rev. Lett.}}
  \textbf{\bibinfo{volume}{72}}, \bibinfo{pages}{3439--3443}
  (\bibinfo{year}{1994}).

\bibitem{giovannetti_sub-heisenberg_2012}
\bibinfo{author}{Giovannetti, V.} \& \bibinfo{author}{Maccone, L.}
\newblock \bibinfo{title}{Sub-heisenberg estimation strategies are
  ineffective}.
\newblock \emph{\bibinfo{journal}{Phys. Rev. Lett.}}
  \textbf{\bibinfo{volume}{108}}, \bibinfo{pages}{210404}
  (\bibinfo{year}{2012}).

\bibitem{hayashi_comparison_2011}
\bibinfo{author}{Hayashi, M.}
\newblock \bibinfo{title}{Comparison between the cramer-rao and the mini-max
  approaches in quantum channel estimation}.
\newblock \emph{\bibinfo{journal}{Commun. Math. Phys.}}
  \textbf{\bibinfo{volume}{304}}, \bibinfo{pages}{689--709}
  (\bibinfo{year}{2011}).

\bibitem{chiribella_quantum_2013}
\bibinfo{author}{Chiribella, G.}, \bibinfo{author}{Yang, Y.} \&
  \bibinfo{author}{Yao, A. C.-C.}
\newblock \bibinfo{title}{Quantum replication at the heisenberg limit}.
\newblock \emph{\bibinfo{journal}{Nat. Commun.}} \textbf{\bibinfo{volume}{4}},
  \bibinfo{pages}{2915} (\bibinfo{year}{2013}).

\bibitem{holevo_probabilistic_1982}
\bibinfo{author}{Holevo, A.~S.}
\newblock \emph{\bibinfo{title}{Probabilistic and Statistical Aspects of
  Quantum Theory}} (\bibinfo{publisher}{North-Holland, Amsterdam},
  \bibinfo{year}{1982}).

\bibitem{kolodynski_phase_2010}
\bibinfo{author}{Ko{\l}ody{\'n}ski, J.} \&
  \bibinfo{author}{Demkowicz-Dobrza{\'n}ski, R.}
\newblock \bibinfo{title}{Phase estimation without \textit{a priori} phase
  knowledge in the presence of loss}.
\newblock \emph{\bibinfo{journal}{Phys. Rev. A}} \textbf{\bibinfo{volume}{82}},
  \bibinfo{pages}{053804} (\bibinfo{year}{2010}).

\end{thebibliography}
\end{document}